\documentstyle[epsfig,aps,twocolumn,amssymb,floats,tighten]{revtex}

\def\Journal#1#2#3#4{{#1} {\bf #2}, #3 (#4)}
\def\NIMR{Nucl. Instr. and Meth. in Phys. Res. A}
\def\NPA{Nucl. Phys. A}
\def\PLB{Phys. Lett.  B}
\def\PRL{Phys. Rev. Lett.}
\def\PRC{Phys. Rev. C}

\def\PR{Phys. Rep.} 
\def\ZPA{Z. Phys. A}
\def\EPJA{Eur. J. Phys. A}

\def\ARNPS{Ann. Rev. Nucl. Part. Sc.}

\def\be{\begin{equation}}
\def\ee{\end{equation}}

\newcommand{\pt}{$p_t^{(0)}$}

\newcommand{\ud}{\mathrm{d}}
\newcommand{\buc}     {$\rm^{1}$}
\newcommand{\bud}     {$\rm^{2}$}
\newcommand{\clt}     {$\rm^{3}$}
\newcommand{\gsi}     {$\rm^{4}$}
\newcommand{\dre}     {$\rm^{5}$}
\newcommand{\hei}     {$\rm^{6}$}
\newcommand{\ite}     {$\rm^{7}$}
\newcommand{\kur}     {$\rm^{8}$}
\newcommand{\kor}     {$\rm^{9}$}
\newcommand{\ire}     {$\rm^{10}$}
\newcommand{\war}     {$\rm^{11}$}
\newcommand{\zag}     {$\rm^{12}$}

\title{Directed flow in Au+Au, Xe+CsI and Ni+Ni collisions
and the nuclear equation of state}

\vspace{0.3cm}
\author{
A.\,Andronic\gsi,
W.\,Reisdorf\gsi,
N.\,Herrmann\hei, 
P.\,Crochet\clt, 
J.P.\,Alard\clt, 
V.\,Barret\clt, Z.\,Basrak\zag, N.\,Bastid\clt, G.\,Berek\bud,
R.\,\v{C}aplar\zag,
A.\,Devismes\gsi, P.\,Dupieux\clt, M.\,D\v{z}elalija\zag,
C.\,Finck\gsi, Z.\,Fodor\bud,
A.\,Gobbi\gsi, Yu.\,Grishkin\ite,
O.N.\,Hartmann\gsi, K.D.\,Hildenbrand\gsi, B.\,Hong\kor,
J.\,Kecskemeti\bud, Y.J.\,Kim\kor,
M.\,Kirejczyk\war, P.\,Koczon\gsi, M.\,Korolija\zag, R.\,Kotte\dre, 
T.\,Kress\gsi,
A.\,Lebedev\ite, Y.\,Leifels\gsi, X.\,Lopez\clt, 
M.\,Merschmeyer\hei,
W.\,Neubert\dre,
D.\,Pelte\hei, M.\,Petrovici\buc,
F.\,Rami\ire,
B.\,de Schauenburg\ire, A.\,Sch\"uttauf\gsi,
Z.\,Seres\bud,
B.\,Sikora\war, K.S.\,Sim\kor, V.\,Simion\buc, K.\,Siwek-Wilczy\'nska\war, 
V.\,Smolyankin\ite, M.R.\,Stockmeier\hei, G.\,Stoicea\buc, 
Z.\,Tyminski\gsi$^,$\war,
P.\,Wagner\ire, 
K.\,Wi\'{s}niewski\war, D.\,Wohlfarth\dre,
I.\,Yushmanov\kur,
A.\,Zhilin\ite \\~
(FOPI Collaboration)
} 
\vspace{0.25cm}
\address{
\buc~National Institute for Physics and Nuclear Engineering, Bucharest, 
Romania\\
\bud~KFKI Research Institute for Particle and Nuclear Physics, Budapest, 
Hungary\\
\clt~Laboratoire de Physique Corpusculaire, IN2P3/CNRS,
and Universit\'{e} Blaise Pascal, Clermont-Ferrand, France\\
\gsi~Gesellschaft f\"ur Schwerionenforschung, Darmstadt, Germany\\
\dre~Forschungszentrum Rossendorf, Dresden, Germany\\
\hei~Physikalisches Institut der Universit\"at Heidelberg, Heidelberg, 
Germany\\
\ite~Institute for Theoretical and Experimental Physics, Moscow, Russia\\
\kur~Kurchatov Institute, Moscow, Russia \\
\kor~Korea University, Seoul, South Korea\\
\ire~Institut de Recherches Subatomiques, IN2P3-CNRS, Universit\'e
Louis Pasteur, Strasbourg, France \\
\war~Institute of Experimental Physics, Warsaw University, Poland\\
\zag~Rudjer Boskovic Institute, Zagreb, Croatia\\ 
} 

\begin{document}

\maketitle

\begin{abstract}
We present new experimental data on directed flow in collisions of Au+Au, 
Xe+CsI and Ni+Ni at incident energies from 90 to 400$A$~MeV.
We study the centrality and system dependence of integral and differential 
directed flow for particles selected according to charge.
All the features of the experimental data are compared with Isospin Quantum 
Molecular Dynamics (IQMD) model calculations in an attempt to extract 
information about the nuclear matter equation of state (EoS).
We show that the combination of rapidity and transverse momentum analysis 
of directed flow allow to disentangle various parametrizations
in the model.
At 400$A$~MeV, a soft EoS with momentum dependent interactions is best suited 
to explain the experimental data in Au+Au and Xe+CsI, but in case of Ni+Ni 
the model underpredicts flow for any EoS.
At 90$A$~MeV incident beam energy,
none of the IQMD parametrizations studied here is able to consistently explain 
the experimental data.

\end{abstract}

\vspace{2mm}

\section{Introduction}
The study of collective flow in relativistic heavy-ion collisions has been 
an intense field of research for the past twenty years
(see Refs.~\cite{rei97,her99} for recent reviews).
The ultimate motivation for the whole endeavour has been the extraction of 
the equation of state (EoS) of nuclear matter (see Ref.~\cite{sto86} for an 
early account and Ref.~\cite{dan01} for more recent ones). 
Moreover, the study of highly complex (quantum) many-body dynamics of 
heavy-ion collisions is in itself a challenging task.

The (in-plane) directed (or sideward) flow was predicted for semi-central
heavy-ion collisions on the basis of fluid dynamical calculations \cite{sto80} 
and observed in experiments soon after \cite{gus84,ren84}.
The study of the average in-plane transverse momentum, $\langle p_x\rangle$, 
as a function of rapidity, $y$, provides an easy and intuitive way
of quantizing the directed flow \cite{dan85}.
For beam energy range up to a few GeV per nucleon, the experimental 
\cite{dan85,dos86,dos87,kea88,ogi89,zha90,wes93,ram95,par95,hua96,cha97,cro97a,cro97b,pak97,ram99,mag00,liu00} 
and theoretical \cite{kap81,mol84,aic87,bon87,pei89,koc90,gal90,bla91,aic91,lan91,jae92,pan93,ono93,ins94,sof95,fuc96,har98,sah98,ins00}
studies of this phenomenon have provided important understanding of its 
features.
Experimentally, the dependences of directed flow on 
incident energy \cite{dos86,par95,cro97a,cha97,ram99}, 
centrality \cite{dos86,ram95,hua96,cro97a,cro97b,pak97,ram99},
particle type \cite{dos87,par95,hua96,cro97a,cro97b,pak97,ram99},
system size \cite{dos86,cha97,ram99} and isospin \cite{pak97} 
have been determined.
These main features have been quantitatively reproduced by microscopic 
transport models of either Quantum Molecular Dynamics (QMD)
\cite{pei89,aic91,ram95,par95,cro97b,har98,liu00} 
or Boltzmann-Uehling-Uhlenbeck (BUU)
\cite{kea88,koc90,gal90,pan93,ins94,hua96,fuc96,sah98,ins00,liu00}
type.
Despite all these efforts, a definite conclusion on the EoS has not yet 
been achieved. 
As pointed out early on \cite{aic87,gal90,bla91}, the momentum 
dependent interactions (MDI) play a crucial role in the determination 
of the EoS.
The directed flow is influenced in addition by the (in-medium) 
nucleon-nucleon cross section ($\sigma_{nn}$) \cite{pan93,ono93,li99}. 
Moreover, consistency is needed in deriving EoS together with both MDI 
and $\sigma_{nn}$ \cite{jae92,ins00}.
The importance to combine various observables, each sensitive (ideally) 
to one particular aspect of the parameters that influence EoS was emphasized 
recently \cite{dan01}.

The results quoted above have been obtained from the analysis of integrated 
directed flow.
The first analysis of differential directed flow (DDF) was introduced by 
Pan and Danielewicz \cite{pan93}, who studied the transverse momentum 
($p_t$) dependence of the first order Fourier coefficient, 
$ v_1=\langle\cos(\phi)\rangle$, 
where $\phi$ is the angle with respect to the reaction plane.
The DDF was studied around the balance energy ($E_{bal}$, which is the 
energy of disappearance of flow \cite{wes93,par95,mag00})
by Li and Sustich \cite{li99}, who unraveled its interesting patterns.
They also pointed out the marked sensitivity of DDF to both EoS and 
nucleon-nucleon cross section ($\sigma_{nn}$).
At AGS energies the DDF was studied both experimentally \cite{bar99} and 
theoretically \cite{li96,vol97}.
Recently, we have completed the first experimental analysis of DDF for Au+Au
collisions at incident energies from 90 to 400$A$~MeV \cite{and01}.
We have found interesting patterns of the differential flow, evolving as
a function of incident energy, particle type and rapidity.
In particular, the study of high-$p_t$ particles is important because,
as proposed in Ref.~\cite{gai01}, they are good messengers 
from the high density state of the collision.
The DDF could additionally provide snapshots of the flow development 
during the time of the collision.

In this paper we present new experimental data on directed flow in collisions 
of Au+Au, Xe+CsI and Ni+Ni.  
The complete coverage of the FOPI detector makes possible precision studies
of flow, refining our earlier studies done with Phase I data 
\cite{ram95,cro97a,cro97b,ram99}.
Following our recent exploration of the energy range of 90 to 400$A$~MeV
for Au+Au \cite{and01}, we focus here on the centrality (for Au+Au) and
on system size dependence of directed flow. This analysis has been performed
for the incident energies of 250 and 400 $A$~MeV, for particles with $Z$=1 
and $Z$=2. 
To avoid overloading, only a selection of results is included 
in the body of the paper. In the Appendix we provide additional
figures to complete the data set.
After describing the detector, the method of analysis, and the corrections 
applied to data, we study the centrality and system dependence of directed 
flow over the complete forward rapidity range, both in terms of ($p_t$) 
integrated and in a differential way.
All the features of the experimental data are then compared with IQMD model
calculations for the incident energies of 90 and 400$A$~MeV. 

\section{Set-up and data analysis} 

\label{sect-setup}
The data have been measured with a wide phase-space coverage using the 
FOPI detector \cite{gob93} at GSI Darmstadt. 
The reaction products are identified by charge ($Z$) in the forward Plastic 
Wall (PW) at 1.2$^\circ <\theta_{lab}<$~30$^\circ$ using time-of-flight 
(ToF) and specific energy loss. In the Central Drift Chamber 
(CDC), covering 34$^\circ <\theta_{lab}<$~145$^\circ$, the particle 
identification is on mass (mass number $A$), obtained using magnetic 
rigidity and the energy loss. 
For PW the $Z$ resolution is 0.13 charge units for $Z$=1 and 0.14 for 
$Z$=2, while for CDC the mass resolution varies from 0.20 to 0.53 mass 
units for $A$=1 to $A$=4.
The contamination of $Z$=1 in the $Z$=2 sample varies from 6\% to 10\% (from
Ni+Ni at 250$A$~MeV to Au+Au at 400$A$~MeV) for the PW and is up to 20\% for 
the CDC (where it is the contamination of $A$=1,2 and 3 in the $A$=4 sample).
The PW measures the velocity of particles via ToF with an average resolution 
of 150~ps.
For the CDC, the relative momentum resolution $\sigma_{p_{t}}/p_{t}$ varies 
from 4\% for $p_{t}<$~0.5~GeV/c to about 12\% for $p_{t}$=2~GeV/c.
For more details on the detector configuration for this experiment see 
Ref.~\cite{and00}.

\begin{figure}[hbt]
\centering\mbox{\epsfig{file=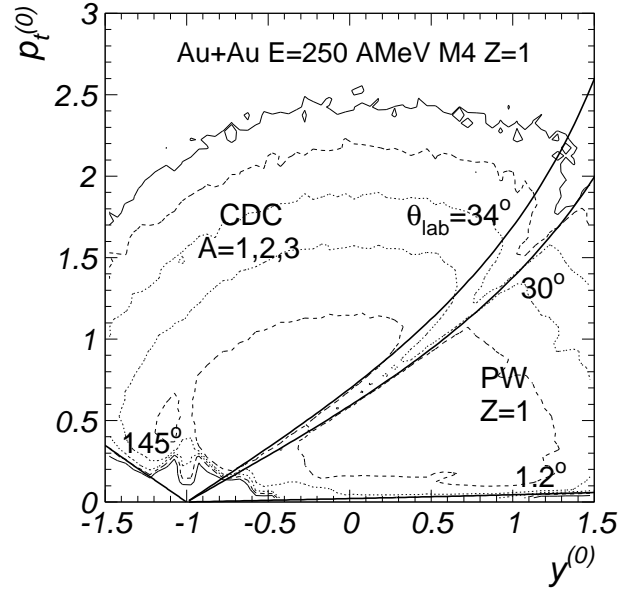, width=0.48\textwidth}}
\caption{FOPI detector acceptance: phase-space distribution for $Z$=1 particles 
measured in the centrality bin M4 of the reaction Au+Au at 250$A$~MeV. 
The intensity contours are spaced logarithmically. The thicker lines mark 
the geometrical acceptance of different subdetectors.} 
\label{fig-x0} 
\end{figure}

The phase-space coverage of the FOPI detector is presented in Fig.~\ref{fig-x0}
for particles with $Z$=1 (for PW) and $A$=1,2,3 (for CDC), measured in 
semi-central collisions Au+Au at incident energy of 250$A$~MeV.
To compare different incident energies and particle species, we use normalized 
center-of-mass (c.m.) transverse momentum (per nucleon) and rapidity, 
defined as
$$
p_t^{(0)}=(p_t/A)/(p_P^{\mathrm{c.m.}}/A_P), 
\quad 
y^{(0)}=(y/y_P)^{\mathrm{c.m.}},
\label{eq-1}
$$
where the subscript $P$ denotes the projectile.
For the PW coverage, shadows around $\theta_{lab}=$~7$^\circ$ and 
~19$^\circ$ are visible, arising from subdetector borders and frames.

For the centrality selection we used the charged particle multiplicities, 
classified into five bins, M1 to M5. 
The variable $E{rat}=\sum_{i}E_{\perp,i}/\sum_{i}E_{\parallel,i}$ 
(the sums run over the transverse and longitudinal c.m. kinetic energy 
components of all the products detected in an event) has been additionally 
used for a better selection of the most central collisions (M5 centrality bin).
The geometric impact parameters interval for the centrality bins
M3, M4, and M5 for Au+Au system at 400$A$~MeV studied here are presented
in Table~\ref{tab-1}.

\begin{table}[hbt]
\caption{The geometric impact parameters intervals $\Delta b_{geo}$ and the 
correction factors for the reaction plane resolution, 
1/$\langle\cos\Delta\phi\rangle$, for three centrality bins of Au+Au collisions
at the incident energy of 400$A$~MeV.}
\label{tab-1}
\begin{tabular}{lccc}
Centrality bin ~~    &  ~~M3 & ~~M4  & ~~M5  \\ \hline
 $\Delta b_{geo}$ (fm)                  & 6.1-7.6 & 1.9-6.1 & 0-1.9 \\
 1/$\langle\cos\Delta\phi\rangle$       & 1.05    & 1.04    & 1.17   \\
\end{tabular}
\end{table}

\begin{table}[hbt]
\caption{The geometric impact parameters intervals $\Delta b_{geo}$,
the reduced impact parameters $\langle b_{geo}\rangle/b_{geo}^{max}$
and the correction factors for the reaction plane resolution, 
1/$\langle\cos\Delta\phi\rangle$, for the three systems at the incident energy 
of 250$A$~MeV, M4 centrality bin.}
\label{tab-2}
\begin{tabular}{lccc}
System ~~                          &  ~~Au+Au    & ~~Xe+CsI  & ~~Ni+Ni  \\ \hline
 $\Delta b_{geo}$ (fm)                  & 1.9-6.1 & 1.7-4.8 & 1.5-3.4 \\
 $\langle b_{geo}\rangle/b_{geo}^{max}$ & 0.31    & 0.29    & 0.27   \\
 1/$\langle\cos\Delta\phi\rangle$       & 1.05    & 1.09    & 1.27   \\
\end{tabular}
\end{table}

 The impact parameter intervals corresponding to the three investigated 
systems at the incident energy of 250$A$~MeV, M4 centrality bin, are presented
in Table~\ref{tab-2} along with the reduced impact parameters
$\langle b_{geo}\rangle/b_{geo}^{max}$. 
$b_{geo}^{max}$ is the maximum geometrical impact parameter, calculated as: 
$b_{geo}^{max}=1.2 (A_P^{1/3}+A_T^{1/3})$ (in fm).
Applying the same recipe for the centrality selection for different systems
the reduced impact parameter is similar, as seen in Table~\ref{tab-2}.

The largest data sample was acquired with what we call ``Medium bias''
trigger, which accepts events roughly corresponding to centrality bins 
M3, M4 and M5 for Au+Au collisions. For Xe and Ni systems this trigger 
selection amounts to a bias for the M3 centrality. This is the reason
why M3 is not included in the present paper for these systems.
We have collected ``Minimum bias'' data for all systems, but the statistics 
is far smaller, thus not allowing the type of analysis done in this paper.

\subsection{The reaction plane determination and the correction for 
its resolution}

The reaction plane has been reconstructed event-by-event using the transverse
momentum method \cite{dan85}.
All charged particles detected in an event have been used, except 
a window around midrapidity ($|y^{(0)}|<$ 0.3) to improve the resolution.
The particle-of-interest has been excluded to prevent autocorrelations.

The correction of the extracted values due to the reconstructed reaction 
plane fluctuations has been done using the recipe of Ollitrault \cite{oli97}. 
The resolution of the reaction plane azimuth, $\Delta\phi$, can be extracted
by randomly dividing each event in two subevents and calculating for each one
the reaction plane orientation, $\Phi_1$ and $\Phi_2$ \cite{dan85,oli97}.
From the resolution, quantified as $\langle\cos(\Phi_1-\Phi_2)\rangle$, 
the correction factors, 1/$\langle\cos\Delta\phi\rangle$ can be calculated
\cite{oli97} (see Ref.~\cite{and00} for more technical details).
For the experimental data the correction factors for the centrality bins M3, 
M4, and M5 for Au+Au system at 400$A$~MeV are presented in Table~\ref{tab-1}.
The values for the three investigated systems at the incident energy of 
250$A$~MeV, M4 centrality bin, are presented in Table~\ref{tab-2}.

\begin{figure}[hbt] 
\centering\mbox{\epsfig{file=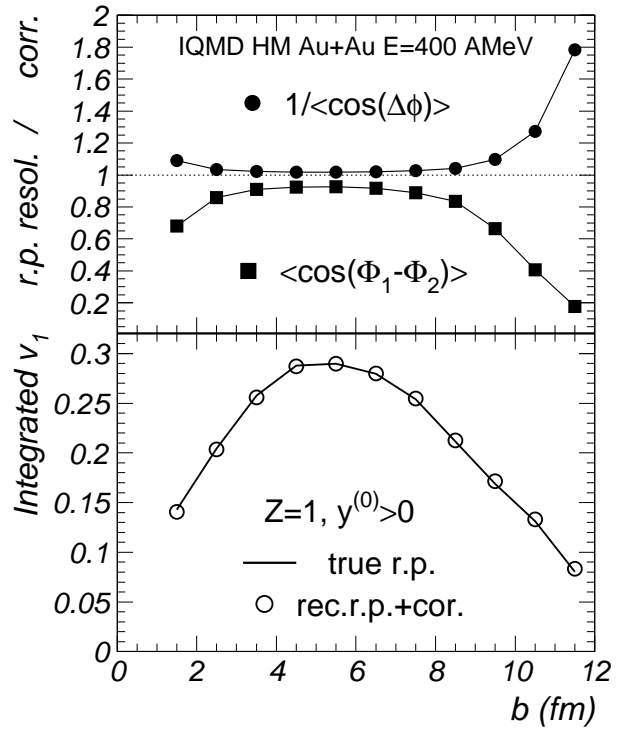, width=.45\textwidth}}
\caption{Upper panel: the resolution of the reconstructed reaction plane 
(squares) and the corresponding correction factors (dots).
Lower panel: $v_1$ values for the true (continuous line) and reconstructed and
corrected (dashed line) reaction plane. IQMD HM events were used for these 
studies.} \label{fig-x3}
\end{figure} 

The accuracy of the reaction plane correction procedure was checked using 
Isospin Quantum Molecular Dynamics (IQMD) \cite{aic91} events analyzed in 
the same way as the data. 
The results are shown in Fig.~\ref{fig-x3} for IQMD events at 400$A$~MeV.
The upper panel presents the resolution and the correction factor as a 
function of the impact parameter, $b$. Their dependence on centrality reflects
mainly the dependence of the strength of the directed flow (see lower panel), 
but also a finite number effect \cite{dan85}, evident towards peripheral 
collisions.
The lower panel of Fig.~\ref{fig-x3} presents the centrality dependence of 
integrated $v_1$ values for $Z$=1 particles (for the forward hemisphere), 
derived from IQMD events in two cases: i) with respect to the true reaction 
plane (known in the model) and ii) with respect to the reconstructed reaction 
plane and using the correction according to \cite{oli97} (correction factors 
of the upper panel of Fig.~\ref{fig-x3}). The agreement between the two cases 
is perfect, down to most peripheral collisions, where correction factors up to 
2 are necessary.

Alternative methods of flow analysis have been proposed recently \cite{bor02}.
However, because, for our energy domain, the flow of nucleons is at its 
maximum and produced particles are very rare,
the impact of these refined methods is expected to be minor.

\subsection{The influence of FOPI detector on the flow measurements}
\label{sect-cor}

As seen in Fig.~\ref{fig-x0}, the complete phase space coverage of the FOPI 
detector (in its Phase II) is hampered by one empty region, 
corresponding to polar angles $\theta_{lab}$ from 30$^\circ$ to 34$^\circ$ 
in the laboratory frame.
Additional detector shadows around 7$^\circ$ and 19$^\circ$ are present too.
We have studied the effect of the FOPI acceptance using IQMD transport model 
\cite{aic91}.
The IQMD events were analyzed in the same way as the experimental data.

\begin{figure}[hbt] 
\centering\mbox{\epsfig{file=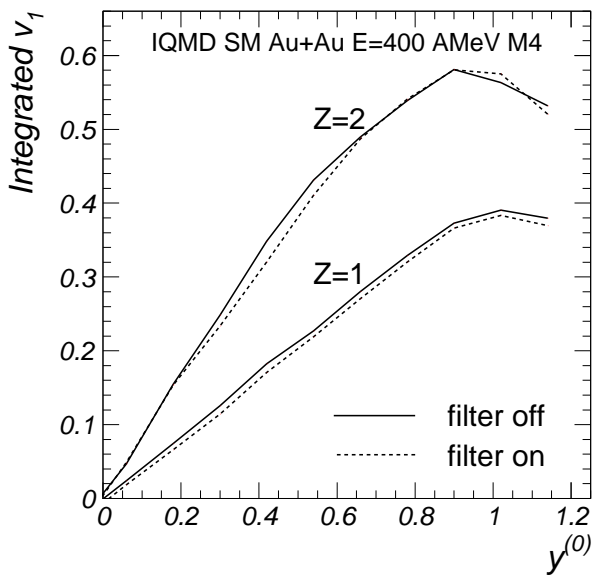, width=.47\textwidth}}
\caption{The effect of the FOPI detector filter on $v_1$ values for $Z$=1 and 
$Z$=2 particles for the incident energy of 400$A$~MeV, M4 centrality bin. 
IQMD SM events were used for this comparison.} 
\label{fig-x1} 
\end{figure}

The results are presented in Fig.~\ref{fig-x1}, where the integrated $v_1$
values as a function of rapidity for $Z$=1 and $Z$=2 particles are shown for
the ideal case of total coverage (full lines) and when the FOPI filter
is employed (dashed lines).
IQMD SM events for the incident energy of 400$A$~MeV, M4 centrality bin 
were used for this comparison.
With the present configuration of the detector, the measured $v_1$ values are 
very close to the ideal case.
We note that, for our published directed flow data 
\cite{ram95,cro97a,cro97b,ram99}, although the effect of the FOPI 
acceptance on the directed flow results was quite small \cite{ram95,cro96},
its magnitude was comparable to the difference between soft and hard EoS.

\begin{figure}[hbt] 
\centering\mbox{\epsfig{file=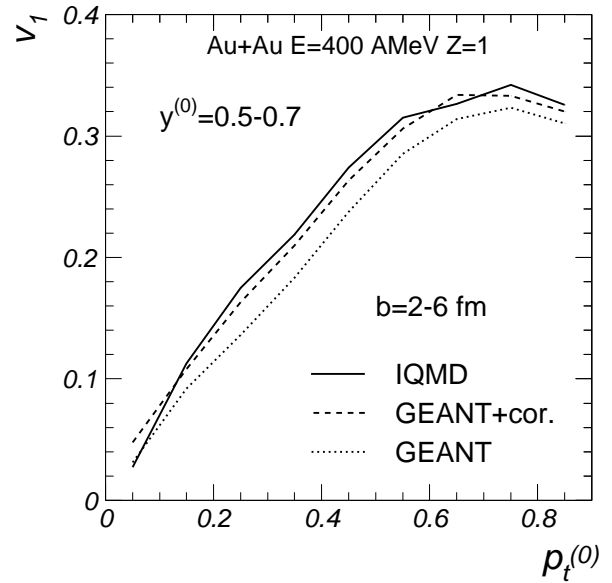, width=.47\textwidth}}
\caption{The effect of the FOPI detector on $v_1$ values for $Z$=1 particles, 
for the incident energy of 400$A$ MeV, M4 centrality bin.
IQMD events were used for this comparison. Three cases are compared:
the simple geometric FOPI filter (full line) and GEANT simulation without 
(dotted line) and with (dashed line) the multiple hit correction. } 
\label{fig-x2} 
\end{figure}

An important ingredient in the present analysis is the correction for 
distortions due to multiple hit losses. 
As an example, in case of PW, despite its good granularity (512 independent 
modules \cite{gob93}), average multiple hit probabilities of up to about 
9\% at 400$A$~MeV are registered for the multiplicity bin M4. 
Because of the directed flow, the average number of particles detected over 
the full PW subdetector is up to 2 times higher in-plane than out of the 
reaction plane.
As a consequence, the losses due to multiple hit are strongly correlated
with the directed flow and follow its dependences on incident energy,
centrality and system size.
These losses lead to an underestimation of the measured directed flow and 
need to be taken into account. 
We developed a correction procedure based on the experimental data, 
by exploiting the DDF left-right symmetry with respect to midrapidity. 
The correction acts upon $v_1$ values (namely only on average values) and 
is deduced for each system, energy and centrality separately.
Due to the flow profile in the polar angle, it depends also on transverse 
momentum. It is larger for $Z$=2 compared to $Z$=1 particles. 
The correction was derived in a window around midrapidity ($|y^{(0)}|<0.1$)
and propagated for other rapidity windows for each \pt bin along the lines 
of constant $\theta_{lab}$ to follow the detector segmentation.
It reaches up to 12\% for Au+Au at 400$A$~MeV and is almost negligible at 
90$A$~MeV. 
At 400$A$~MeV it is up to 5\% for Xe+CsI and up to 2\% for Ni+Ni.

The procedure was checked and validated using IQMD events passed through 
a complete GEANT \cite{geant} simulation of the detector.
IQMD events were used for this study, at the incident energy of 400$A$~MeV.
The results are presented in Fig.~\ref{fig-x2} for $Z$=1 particles in M4 
centrality bin.
The $v_1$ values extracted from IQMD events as inputs into a complete GEANT 
simulation of the detector (and analyzed in exactly the same way as the data),
without (dotted line) and with (dashed line) the multiple hit correction,
are compared with the true $v_1$ values, obtained from standard IQMD events 
(actually the same events used for the GEANT simulation) with only a simple 
geometric FOPI filter (full line).
It is obvious that the correction is restoring the ``true'' $v_1$ values 
(IQMD) from the ``measured'' GEANT values. Also, the correction used for 
the data is quantitatively reproduced by these simulations. Note that, 
because of larger multiplicities for $Z$=1 particles from IQMD (see Section
\ref{sect-iqmd}), the correction is larger in the simulations than in the data.
As a result of these studies, all the experimental data (both differential
and integrated) have been corrected according to the procedure described
above.

The only source of systematic error on our measured $v_1$ values could be 
the correction for multiple hit losses outlined above. 
However, as we have demonstrated based on complete GEANT simulations, 
this correction is well understood.
As a result, the systematic error depends on incident energy, centrality, 
particle type and \pt. It is below 5\% on the differential $v_1$.
There are exceptions for some points, for which the systematic error arises 
from (rapidity-dependent) regions in \pt in which detector shadows are 
influencing the data.
For those particular points the systematic error is already included in 
the plots.
For the integrated $v_1$ values the error is slightly smaller, up to 4\%,
including the influence of the uncovered region of $\theta_{lab}=30^\circ -
34^\circ$. These values do not include the effect of particle misidentification.

\section{General features of the data}
\label{sect:aa_gen}

\subsection{Centrality dependence}

By varying the centrality of the collision one aims at controlling both the
size of the participant fireball (and consequently the magnitude of the
achieved compression and subsequent expansion) and the size of the spectator
fragment region. 
While semi-central collisions could provide information preferentially on 
(density dependent) EoS, more peripheral reactions can help in pinning down 
the MDI \cite{dan01}.

\begin{figure}[hbt]
\centering\mbox{\epsfig{file=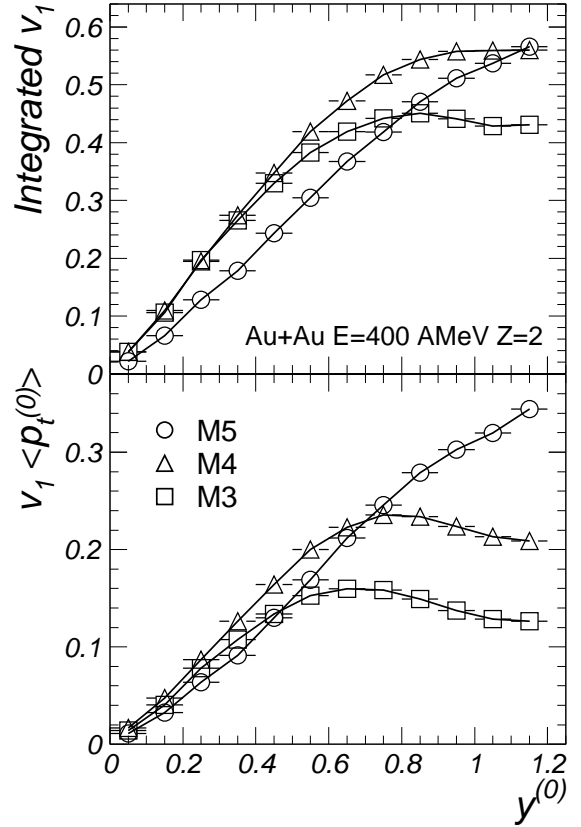,width=.45\textwidth}}
\caption{Centrality dependence of the integrated directed flow as a function 
of rapidity for Au+Au at 400$A$~MeV for $Z$=2 particles. The lines are joining 
the symbols to guide the eye.}
\label{fig-c2} 
\end{figure}

Figure~\ref{fig-c2} shows the centrality dependence of the directed flow
for Au+Au at 400$A$~MeV for $Z$=2 particles. 
Plotted as a function of rapidity are the ($p_t$) integrated $v_1$ values
(upper panel) and those integrated values weighted by the average transverse 
momentum, $\langle p_t^{(0)}\rangle$, for the respective rapidity bin.
This weighted $v_1$ quantity is proportional to the average in-plane transverse 
momentum $\langle p_x\rangle$ ($v_1=\langle p_x/p_t\rangle$) and was chosen
due to its convenience in applying the corrections discussed in Section
\ref{sect-cor}.
First, one can notice that the known behaviour of the slope at midrapidity, 
namely the maximum for intermediate impact parameters (M4 bin), is evident only
for the weighted $v_1$ values (lower panel of Fig.~\ref{fig-c2}).
The asymmetries ($v_1$ values, upper panel) are the same around midrapidity 
for M3 and M4 centrality bins.
Second, in both observables, the most significant dependence on centrality is 
taking place in the spectator region (roughly $y^{(0)}>0.5$) and it is more 
pronounced for the weighted $v_1$ values.
Both the asymmetries and the in-plane transverse momentum reflect the 
influence of the participant and of the spectator size controlled by the 
variation of the centrality.
These distributions are inherently a result of the superposition of collective 
and thermal contributions \cite{ram99}. For a given flow magnitude, higher 
temperatures (presumably achieved for more central collision) would translate
into a smaller effective flow.

\begin{figure}[hbt]
\centering\mbox{\epsfig{file=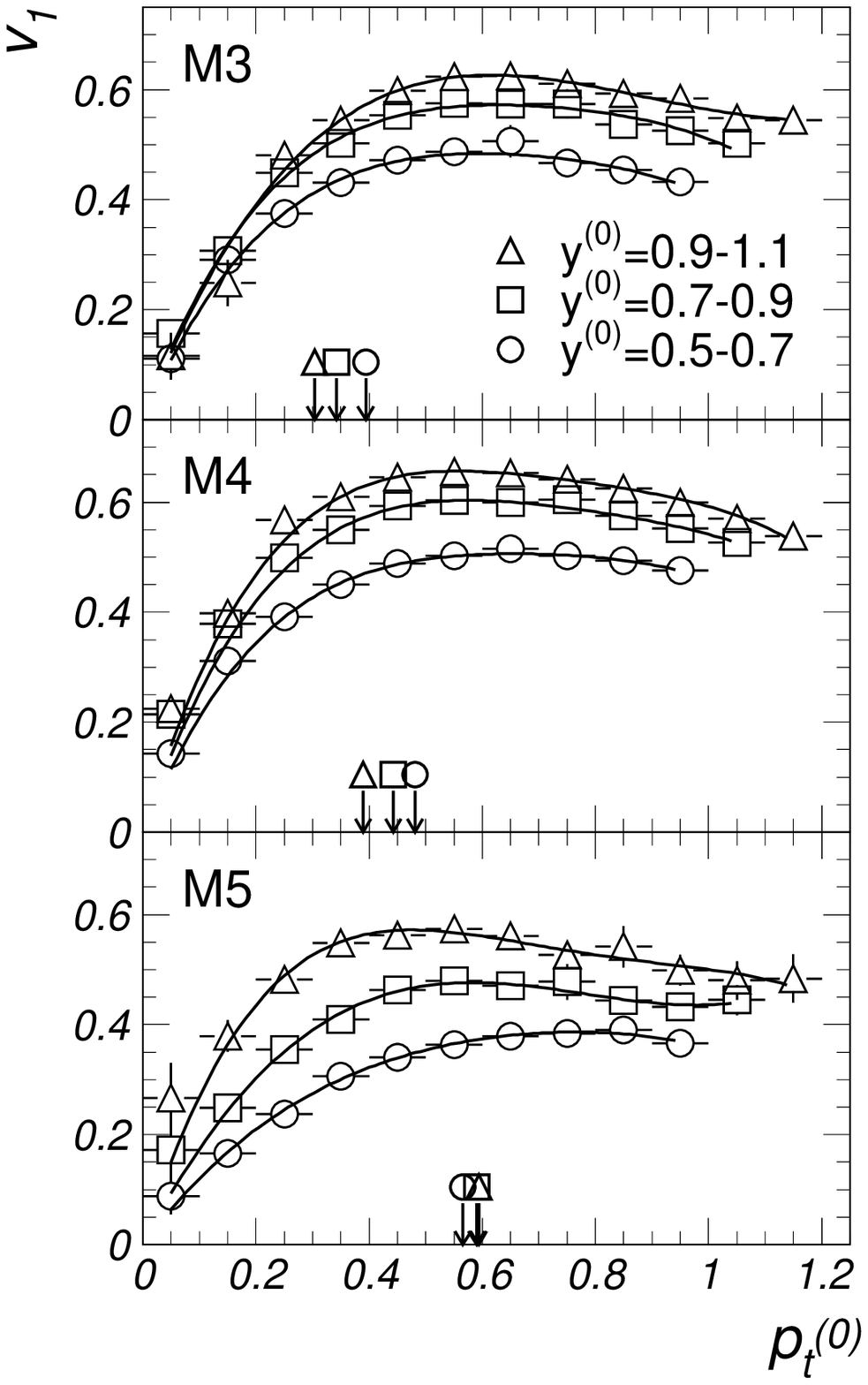,width=.43\textwidth}}
\caption{Differential flow for three centrality bins, in three rapidity 
windows, for $Z$=2 particles for collisions Au+Au at 400$A$~MeV. 
The lines are polynomial fits to guide the eye. The arrows mark the values of 
the average \pt for the corresponding centrality bin.} 
\label{fig-c3} 
\end{figure}

In Fig.~\ref{fig-c3} we show the centrality dependence of the differential 
flow for $Z$=2 particles for Au+Au collisions at 400$A$~MeV.
Three centrality bins (different panels) are compared for three rapidity 
windows (different symbols). 
The lines are polynomial fits to guide the eye.
The rapidity dependence of the DDF in different centrality bins
follows the rapidity dependence of the integrated flow seen in 
Fig.~\ref{fig-c2}: the most pronounced dependence is registered for 
the most central collisions, M5.
As we have already discussed in \cite{and01}, the shape of the DDF (a gradual 
development of a limiting value, followed by a decrease at high \pt) could be
a result of the collision dynamics. 
Part of the high-$p_t$ particles could have been emitted at a pre-equilibrium 
stage, therefore not reaching the maximum compression stage of the reaction.
However, this possibility seems to be ruled out by the observation that the 
high-$p_t$ particles originate preferentially from high-density regions of 
the collision \cite{gai01}.

The arrows in Fig.~\ref{fig-c3} mark the values of the average \pt for the 
corresponding centrality bin, according to the symbols. 
For this incident energy of 400$A$~MeV the value of the projectile 
momentum in the c.m. system is 433~MeV/c per nucleon.
Higher values of average transverse momenta are seen for more central 
collisions as a result of a stronger expansion from a bigger and more 
compressed source.
The dependence of $\langle p_t^{(0)}\rangle$ on rapidity is different for
the most central collisions (M5 bin) compared to semi-central ones, for which
smaller transverse momenta are seen towards the projectile rapidity as a result
of the influence of the spectator matter.

\begin{figure}[htb]
\centering\mbox{\epsfig{file=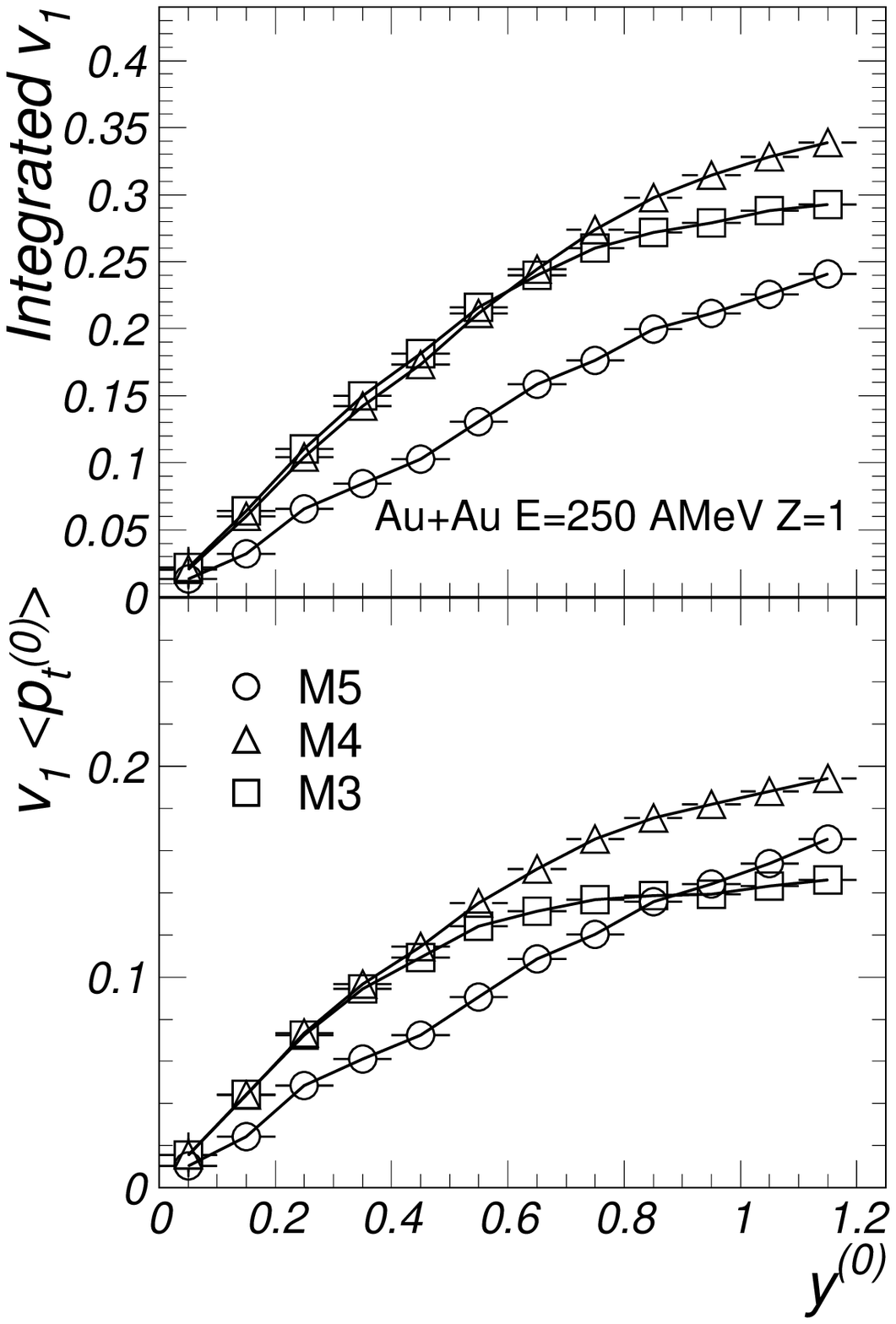, width=.46\textwidth}}
\caption{Centrality dependence of the integrated directed flow as a function of
rapidity for Au+Au at 250$A$~MeV for $Z$=1 particles.}
\label{fig-ad1} 
\end{figure}

\begin{figure}[htb]
\centering\mbox{\epsfig{file=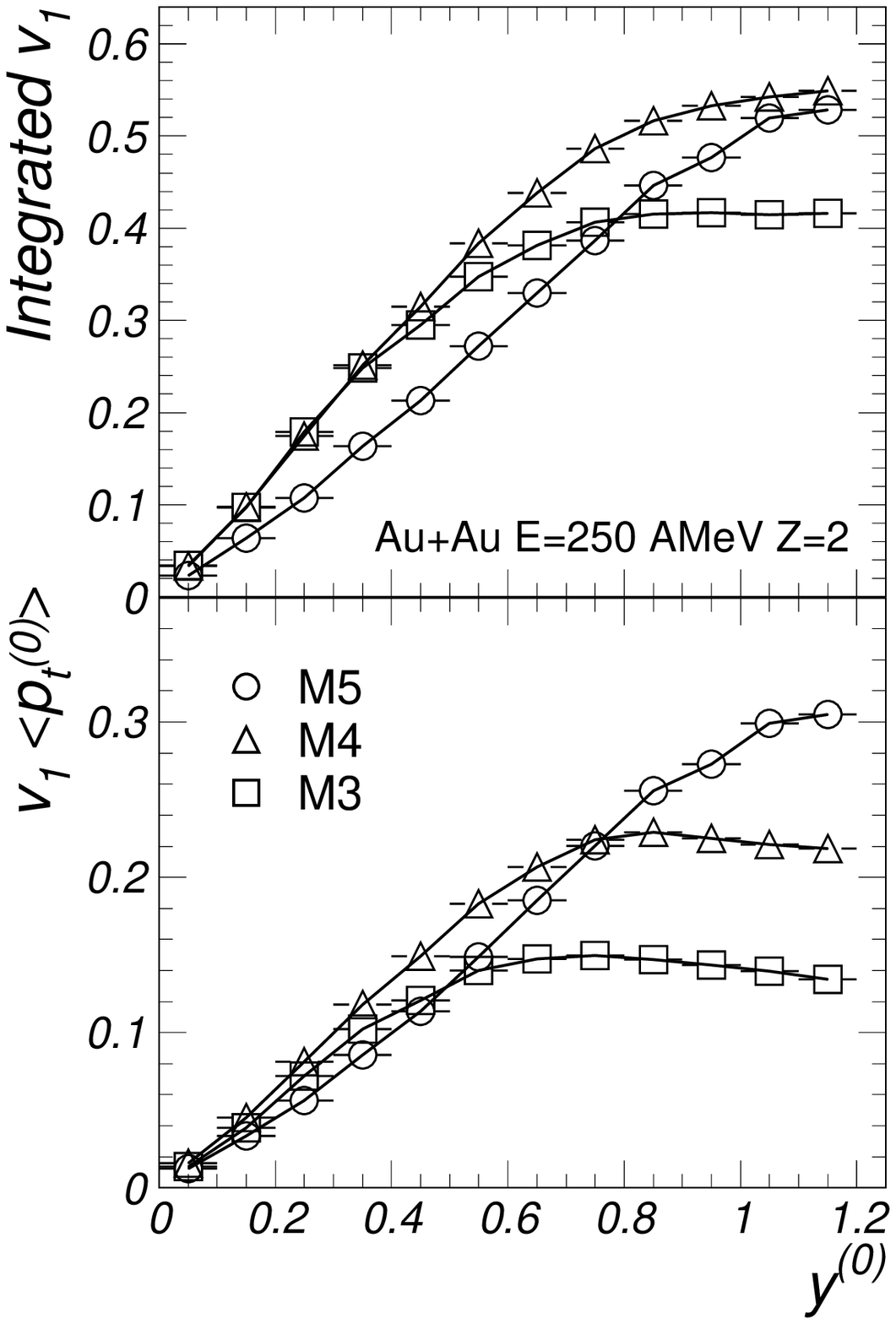, width=.46\textwidth}}
\caption{Centrality dependence of the integrated directed flow as a function of
rapidity for Au+Au at 250$A$~MeV for $Z$=2 particles.}
\label{fig-ad2} 
\end{figure}

\begin{figure}[htb]
\centering\mbox{\epsfig{file=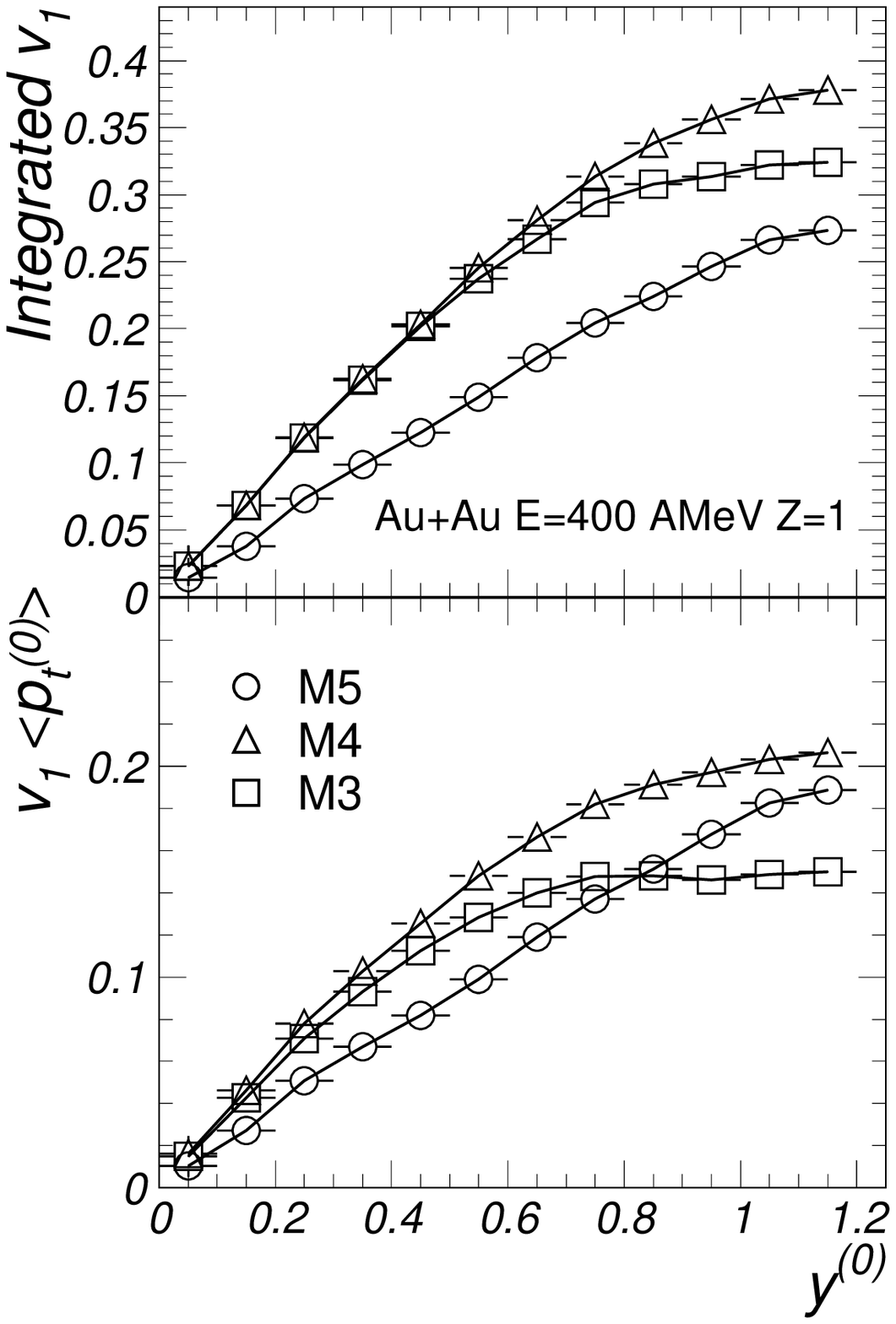, width=.46\textwidth}}
\caption{Centrality dependence of the integrated directed flow as a function of
rapidity for Au+Au at 400$A$~MeV for $Z$=1 particles.}
\label{fig-ad3} 
\end{figure}

\begin{figure}[htb]
\centering\mbox{\epsfig{file=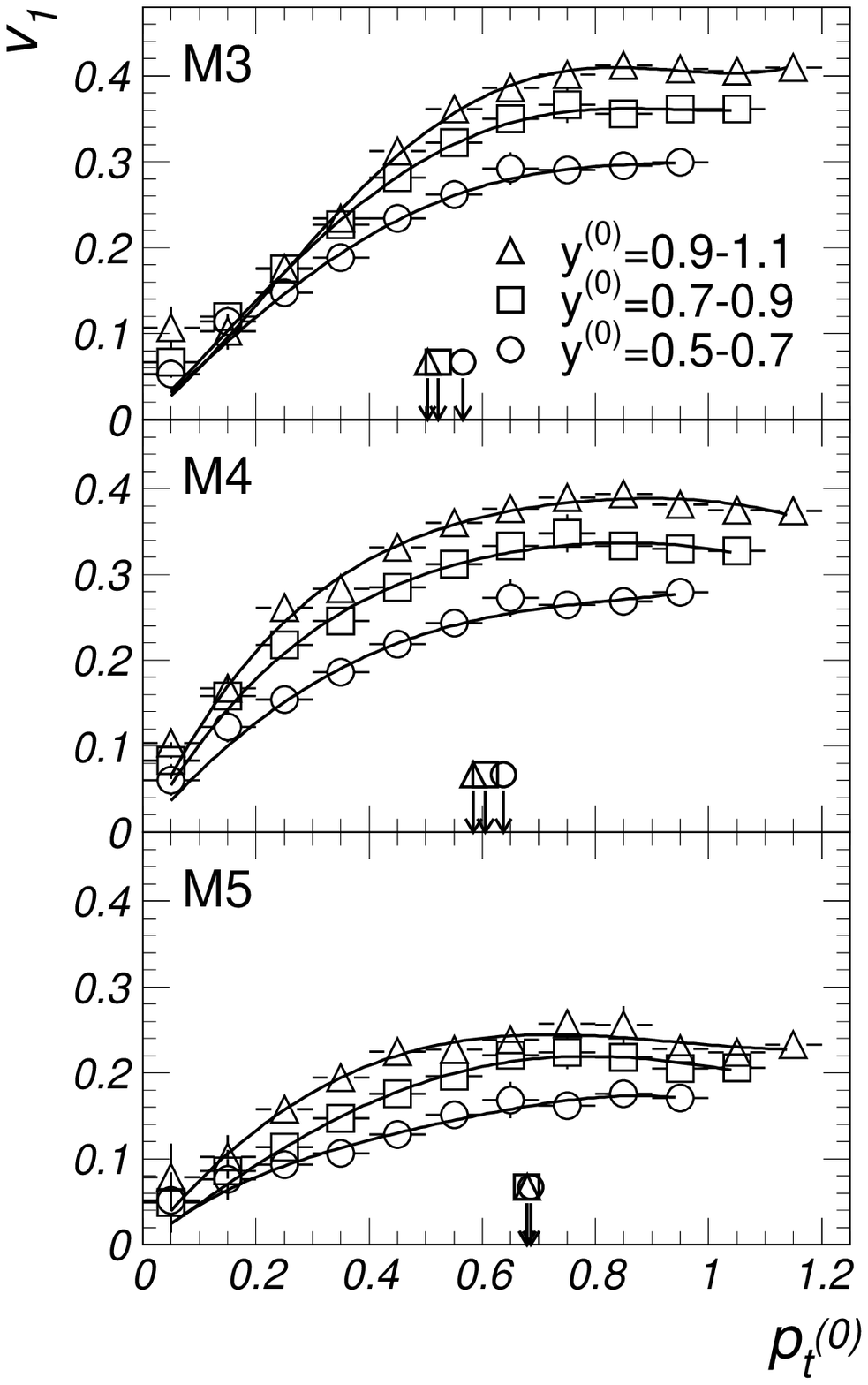, width=.46\textwidth}}
\caption{Differential flow for three centrality bins, in three rapidity 
windows, for $Z$=1 particles for collisions Au+Au at 250$A$~MeV. 
The lines are polynomial fits to guide the eye. The arrows mark the values of 
the average \pt for the corresponding centrality bin.} 
\label{fig-ad4} 
\end{figure}

\begin{figure}[htb]
\centering\mbox{\epsfig{file=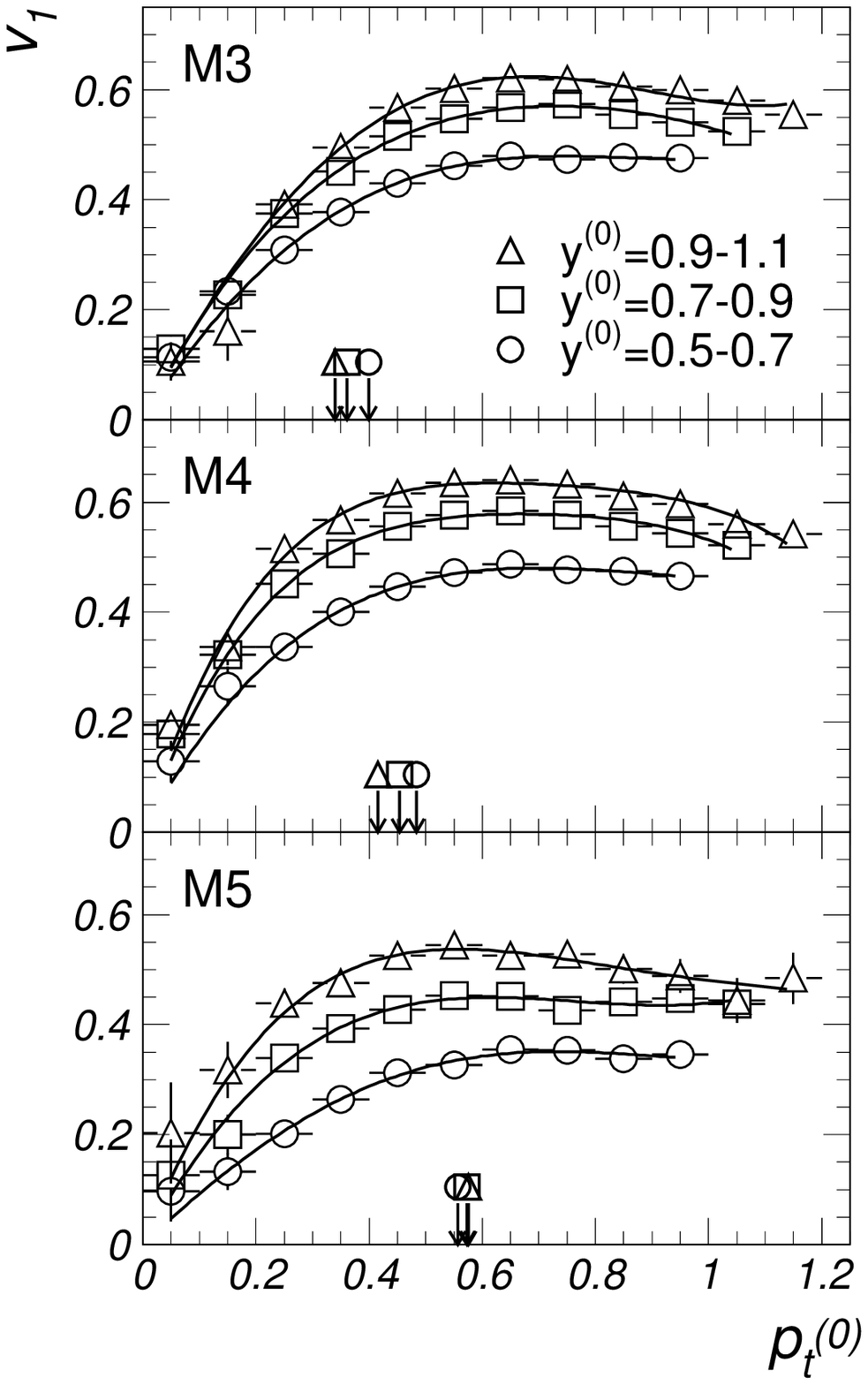, width=.46\textwidth}}
\caption{Differential flow for three centrality bins, in three rapidity 
windows, for $Z$=2 particles for collisions Au+Au at 250$A$~MeV.}
\label{fig-ad5} 
\end{figure}

\begin{figure}[htb]
\centering\mbox{\epsfig{file=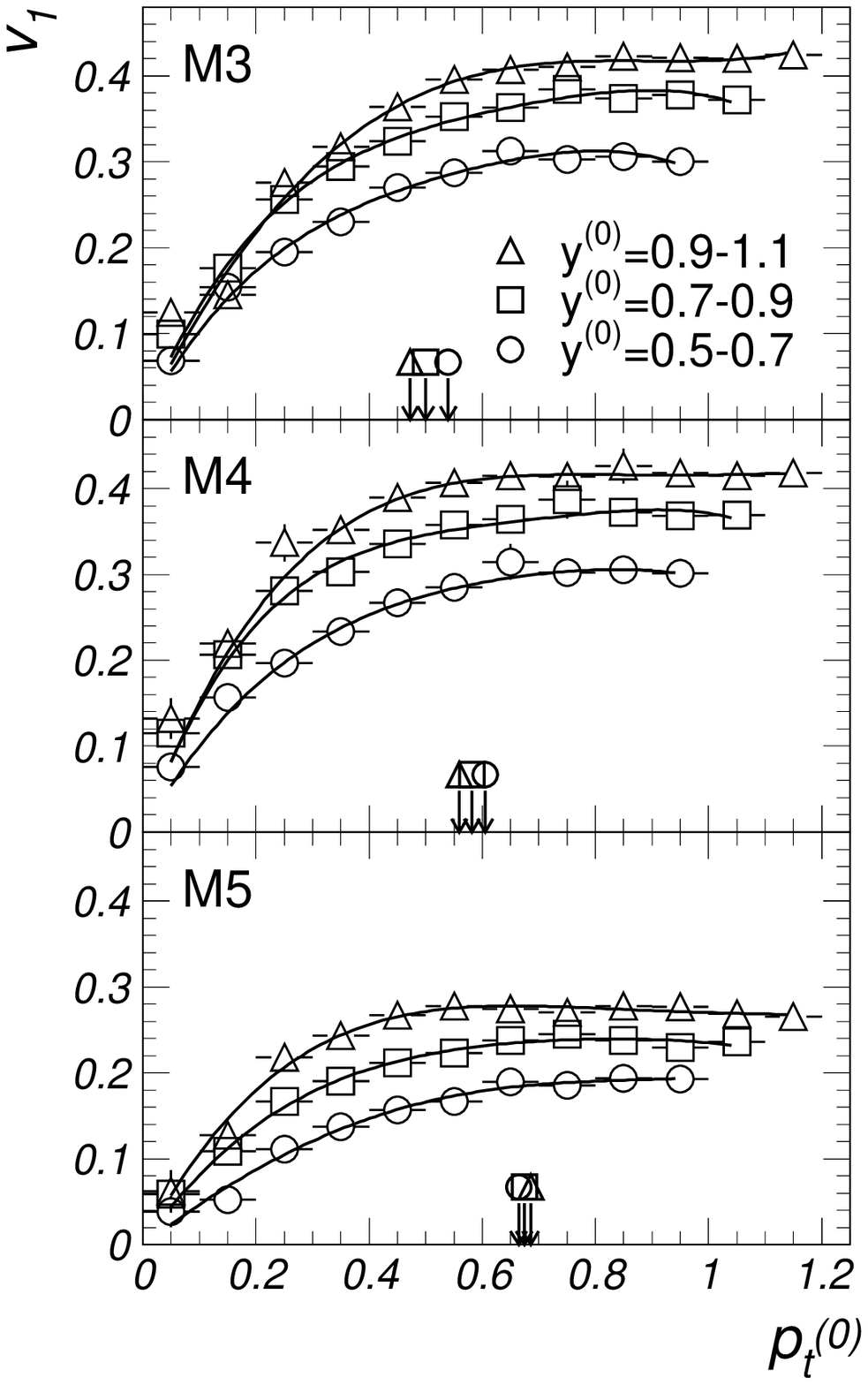, width=.46\textwidth}}
\caption{Differential flow for three centrality bins, in three rapidity 
windows, for $Z$=1 particles for collisions Au+Au at 400$A$~MeV.}
\label{fig-ad6} 
\end{figure}

Similar data as those presented in Fig.~\ref{fig-c2} and \ref{fig-c3} are 
presented in the Appendix for $Z$=1 particles in Au+Au at 400$A$~MeV and 
for $Z$=1 and $Z$=2 particles at 250$A$~MeV (Fig.~\ref{fig-ad1} to 
Fig.~\ref{fig-ad6}).

\subsection{System size dependence}

As for the centrality variation, by varying the system size one aims to 
control the size of both the participant and the spectator. 
However, the question whether transparency plays a role in case of lighter 
systems, needs to be addressed simultaneously, as it results
in a decrease of the achieved compression.
In addition, the surface (or surface-to-volume ratio) can play an important 
role.
It was suggested that the system size dependence of the directed flow could 
give insights about $\sigma_{nn}$ \cite{bla91}.

\begin{figure}[hbt] 
\centering\mbox{\epsfig{file=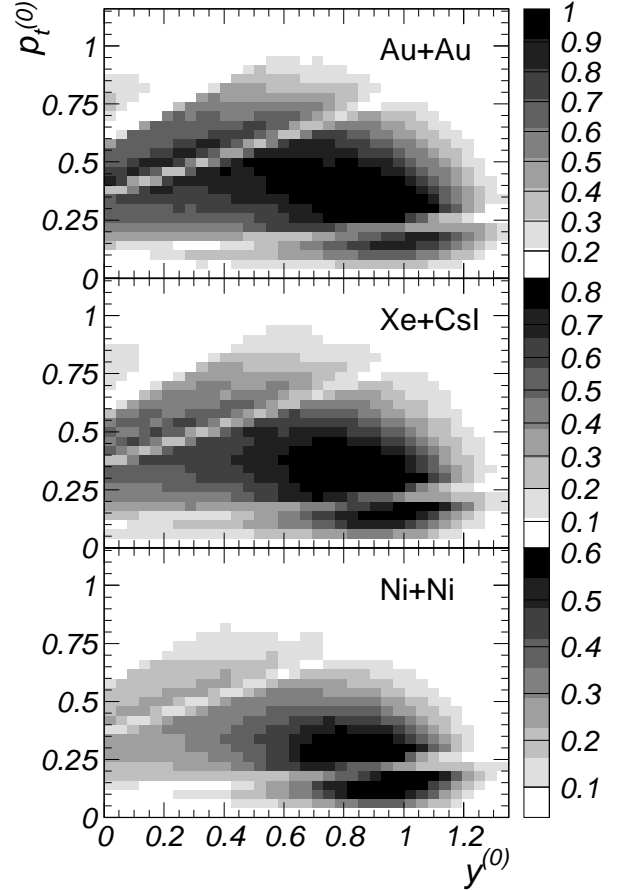, width=.45\textwidth}}
\caption{Phase space distributions 
$\ud^2N/\ud p_t^{(0)}\ud y^{(0)}$ of $Z$=2 particles 
for three systems at the incident energy of 250$A$~MeV, M4 centrality.} 
\label{fig-s1}
\end{figure} 

Figure~\ref{fig-s1} presents the phase space distribution
$\ud^2N/\ud p_t^{(0)}\ud y^{(0)}$ of $Z$=2 particles
for the three systems at the incident energy of 250$A$~MeV, centrality bin M4.
From Au+Au to Ni+Ni system, the phase space population becomes more and more
focused, both in transverse momentum and rapidity. 
This is an indication of the decrease of stopping for lighter systems. 
Maximum density reached in the fireball depends on the system size 
\cite{lan91}, presumably as an effect of different stopping.
On the other hand, due to the sizes of both fireball and spectator, the 
separation between the two regions is clearer (smaller surface contacts) 
for lighter systems.

\begin{figure}[hbt]
\centering\mbox{\epsfig{file=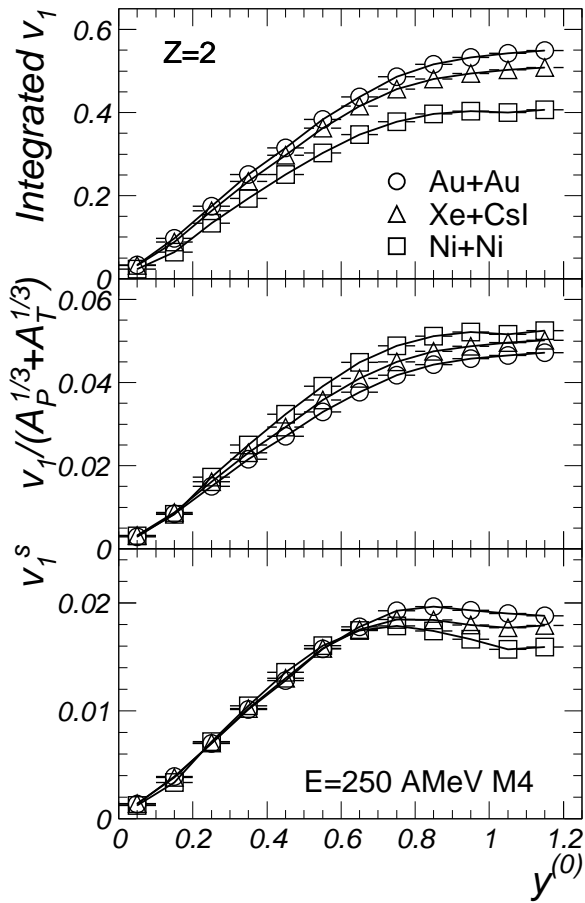,width=.45\textwidth}}
\caption{Integrated directed flow as a function of rapidity for $Z$=2 
particles in the M4 centrality bin of collisions Au+Au, Xe+CsI and Ni+Ni at 
250$A$~MeV. Upper panel: $v_1$ values, middle panel: $v_1$ scaled
by the term $(A_P^{1/3}+A_T^{1/3})$, lower panel: scaled values, $v_1^s$ 
(see text). The lines are joining the symbols to guide the eye.} 
\label{fig-s2}
\end{figure} 

Figure~\ref{fig-s2} shows the system dependence of the directed flow for $Z$=2
particles in the M4 centrality bin of collisions Au+Au, Xe+CsI and Ni+Ni at 
250$A$~MeV. Plotted are the integrated $v_1$ values as a function of rapidity
for three cases: 
i) as such (upper panel);
ii) scaled with the term $(A_P^{1/3}+A_T^{1/3})$, which is proportional to the 
    sum of radii of projectile and target (middle panel);
iii) scaled as: $v_1^s=v_1 \langle p_t^{(0)}\rangle/(A_P^{1/3}+A_T^{1/3}) ,$
where $\langle p_t^{(0)}\rangle$ is the average normalized transverse momentum
for each rapidity bin. 
It is evident that for the first two cases there is no scaling with respect 
to the system size, neither without nor with accounting for the system size 
via the $(A_P^{1/3}+A_T^{1/3})$ term, while the in-plane average transverse 
momenta, proportional to $v_1^s$, shows system size scaling (lower panel
of Fig.~\ref{fig-s2}) for the participant region. 
Somewhat expected, a deviation is present for the spectator part.
We have observed a very similar feature for $Z$=3 particles, while for $Z$=1
the scaling holds over all the forward rapidity domain.

\begin{figure}[hbt]
\centering\mbox{\epsfig{file=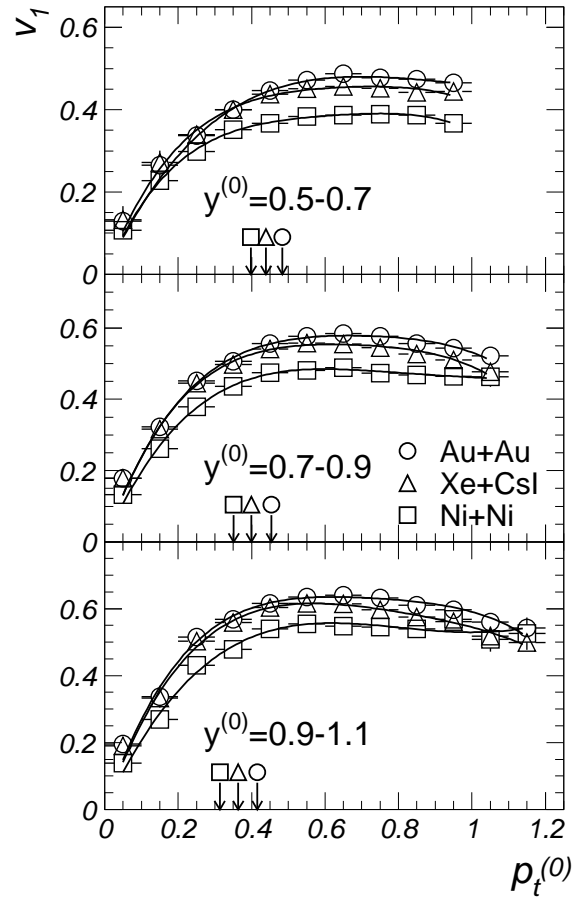,width=.45\textwidth}}
\caption{Differential flow for three systems at 250$A$~MeV, M4 centrality 
bin, for $Z$=2 particles in three windows of rapidity. The lines are polynomial 
fits to guide the eye. The arrows mark the values of the average \pt for the 
corresponding system.}
\label{fig-s3} 
\end{figure}

The $(A_P^{1/3}+A_T^{1/3})$ scaling  has been proposed by Lang et al. 
\cite{lan91}, who, within a BUU model, have found a linear dependence of the 
transverse pressure (leading to transverse momentum transfer) with the reaction 
time (passage time). 
Westfall et al. \cite{wes93} have related the $A^{-1/3}$ dependence 
of $E_{bal}$ to a competition between the attractive mean field (associated 
with the surface, so scaling with $A^{2/3}$) and the repulsive nucleon-nucleon
interaction (scaling as $A$). 
This competition between the two components may be the origin of the
quoted scaling of the transverse pressure with $(A_P^{1/3}+A_T^{1/3})$.
Earlier studies devoted to the slope of $\langle p_x\rangle-y$ distributions
(which translates into a flow angle) have experimentally confirmed such a
scaling \cite{cha97,ram99}.
We note that, in an ideal hydrodynamics the flow angle is a pure 
geometric quantity and does not depend on the system size \cite{sch93}.
The system size dependences presented above may be an interesting effect
of the nuclear forces and/or a consequence of the non-equilibrium nature
of the heavy-ion collisions.

In Fig.~\ref{fig-s3} we present the differential flow for the three systems 
at 250$A$~MeV, M4 centrality bin, for $Z$=2 particles in three windows of 
rapidity (the three panels).
The arrows mark the values of the average \pt for the corresponding centrality 
bin, according to the symbols. 
For this incident energy of 250$A$~MeV the value of the projectile 
momentum in the c.m. system is 342~MeV/c per nucleon.
As seen already in Fig.~\ref{fig-s1}, for all rapidity windows the 
$\langle p_t^{(0)}\rangle$ depend on the system size (again breaking the 
scaling expected from hydrodynamics), suggesting an increase of the 
compression and expansion with the system size.

\begin{figure}[htb]
\centering\mbox{\epsfig{file=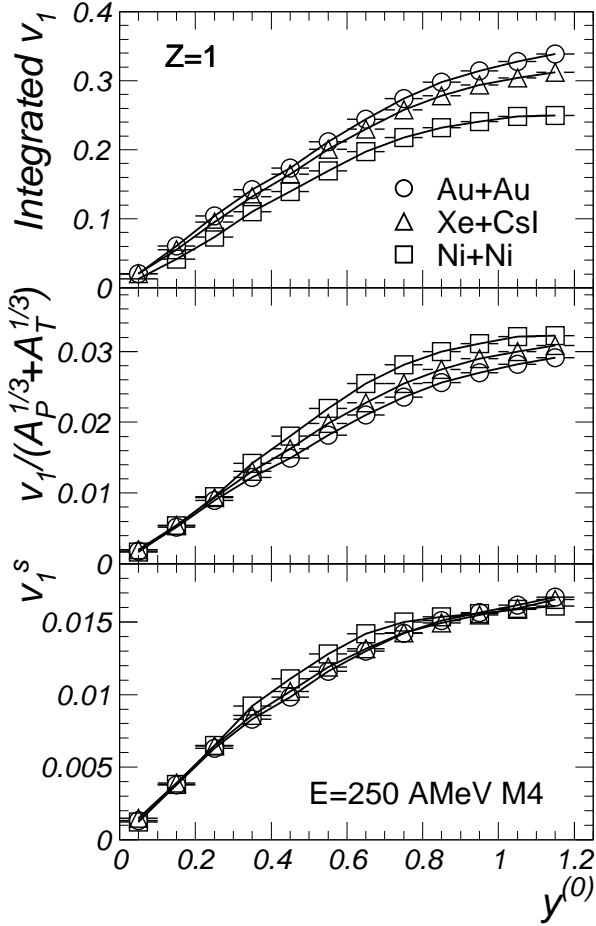, width=.46\textwidth}}
\caption{Integrated directed flow as a function of rapidity for $Z$=1 
particles in the M4 centrality bin of collisions Au+Au, Xe+CsI and Ni+Ni at 
250$A$~MeV. Upper panel: $v_1$ values, middle panel: $v_1$ scaled
by the term $(A_P^{1/3}+A_T^{1/3})$, lower panel: scaled values, 
$v_1^s=v_1 \langle p_t^{(0)}\rangle/(A_P^{1/3}+A_T^{1/3})$.} 
\label{fig-ad7}
\end{figure}

\begin{figure}[htb]
\centering\mbox{\epsfig{file=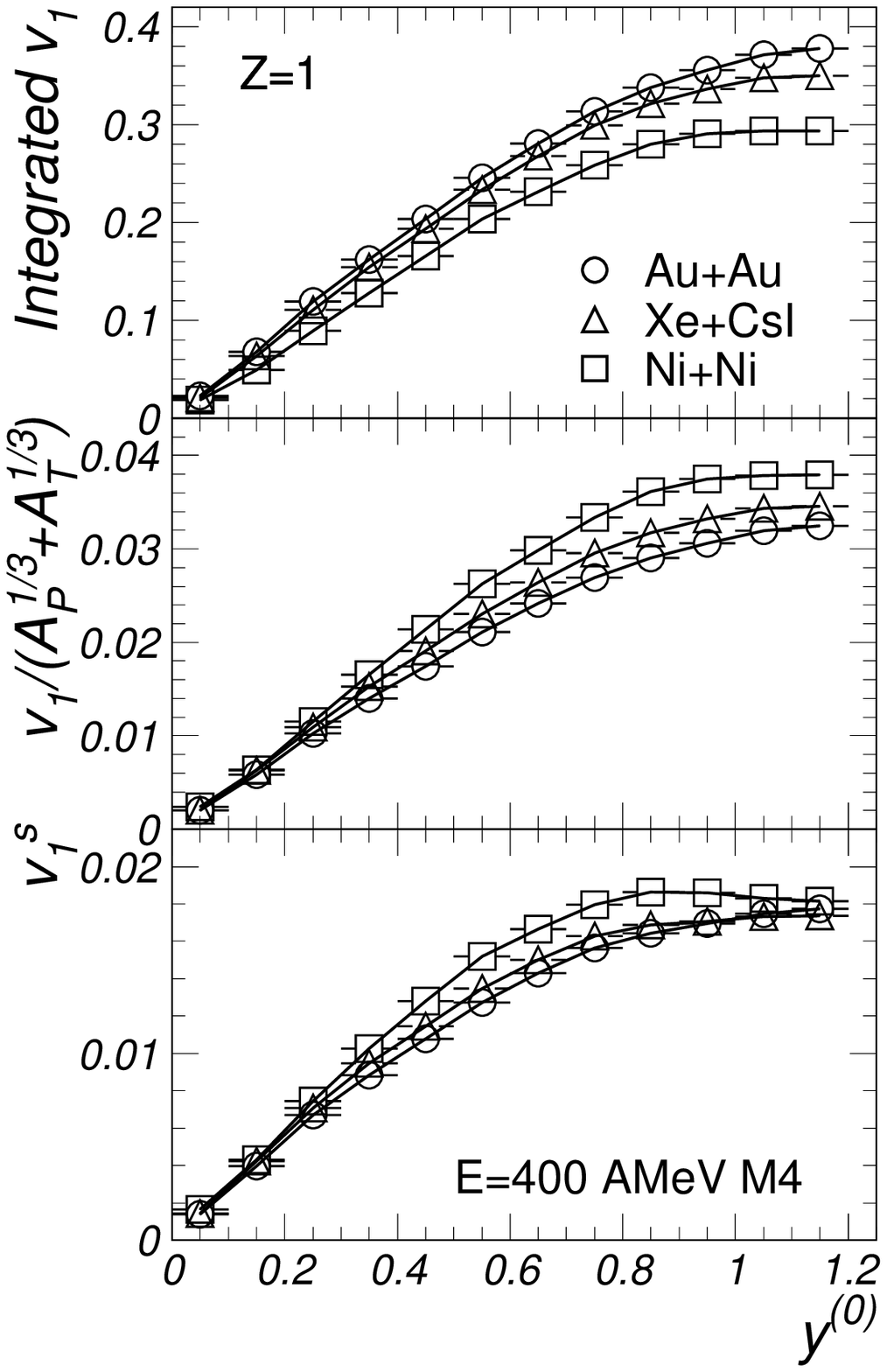, width=.46\textwidth}}
\caption{Integrated directed flow as a function of rapidity for $Z$=1
particles in the M4 centrality bin of collisions Au+Au, Xe+CsI and Ni+Ni at 
400$A$~MeV.} 
\label{fig-ad8}
\end{figure}

\begin{figure}[htb]
\centering\mbox{\epsfig{file=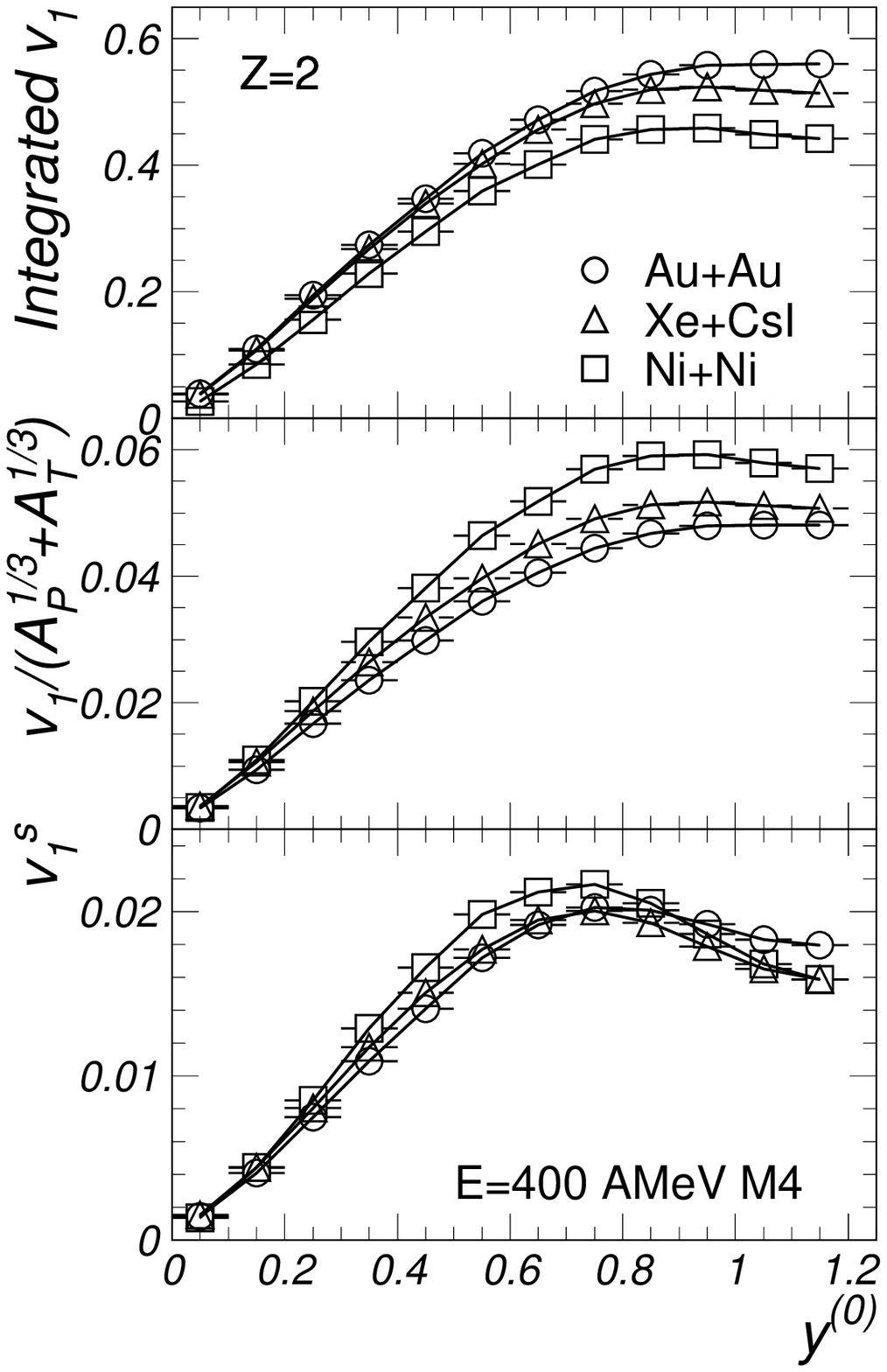, width=.46\textwidth}}
\caption{Integrated directed flow as a function of rapidity for $Z$=2
particles in the M4 centrality bin of collisions Au+Au, Xe+CsI and Ni+Ni at 
400$A$~MeV.} 
\label{fig-ad9}
\end{figure}

\begin{figure}[htb]
\centering\mbox{\epsfig{file=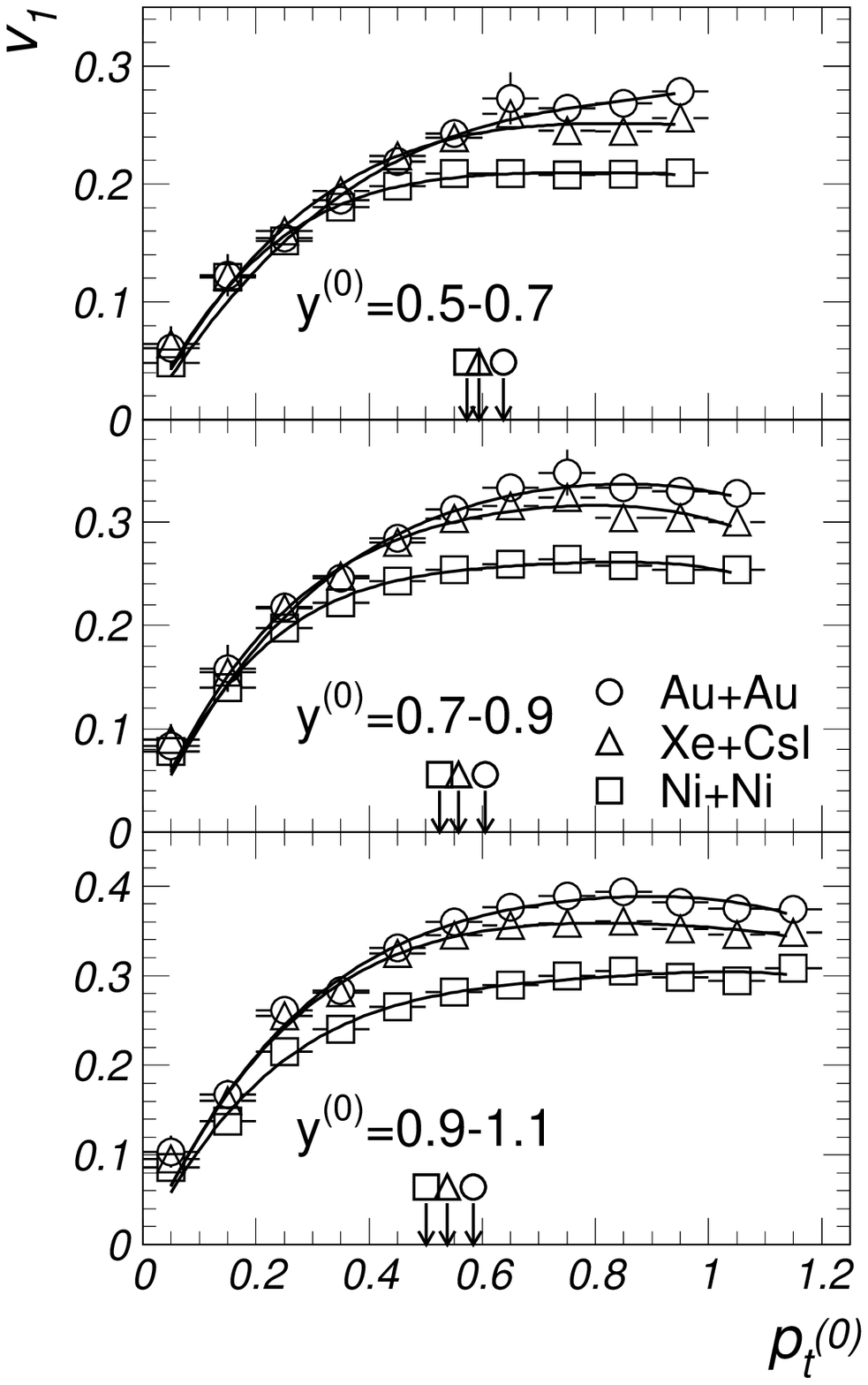, width=.46\textwidth}}
\caption{Differential flow for three systems at 250$A$~MeV, M4 centrality 
bin, for $Z$=1 particles in three windows of rapidity. The lines are polynomial 
fits to guide the eye. The arrows mark the values of the average \pt for the 
corresponding system.}
\label{fig-ad10} 
\end{figure}

\begin{figure}[htb]
\centering\mbox{\epsfig{file=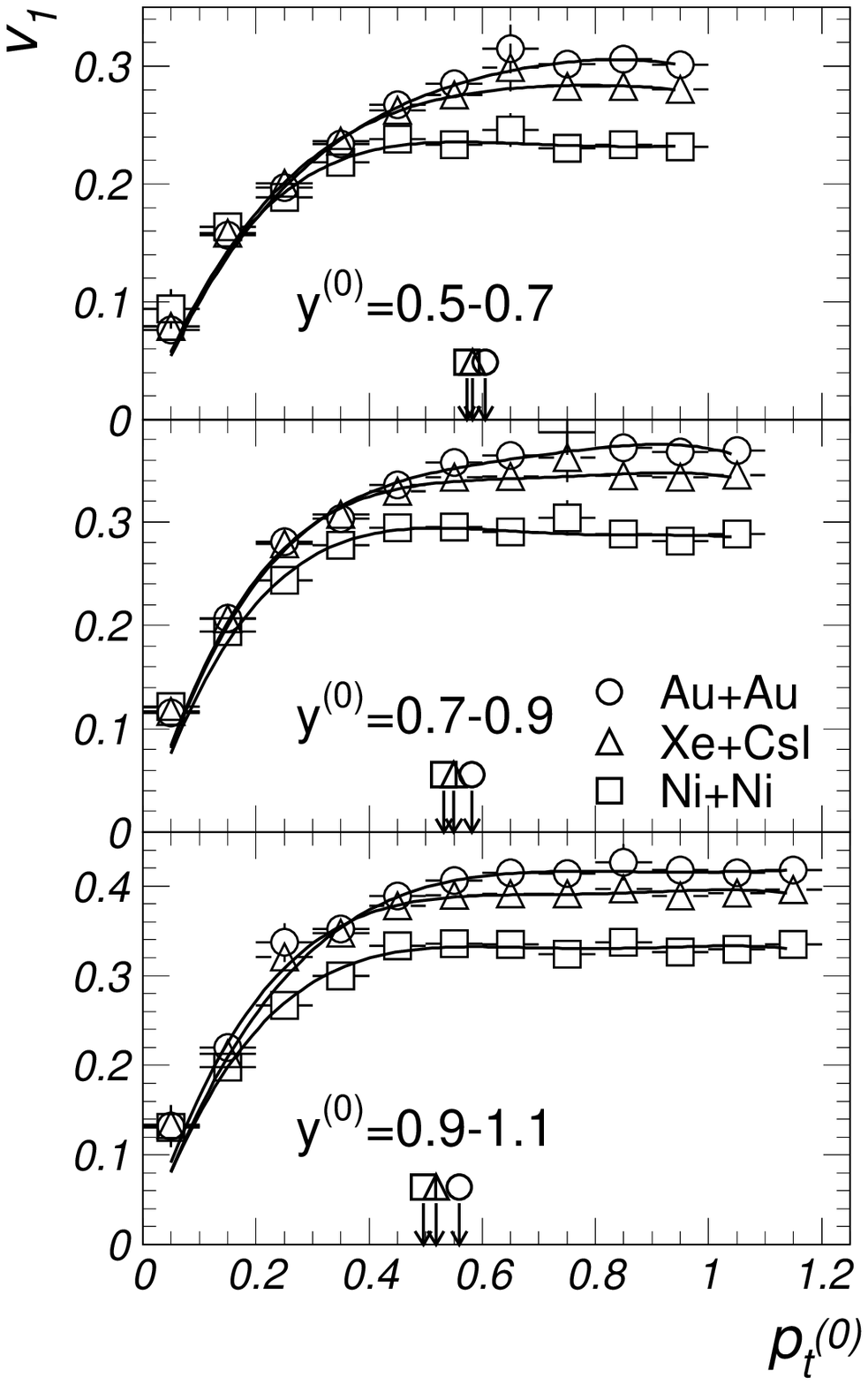, width=.46\textwidth}}
\caption{Differential flow for three systems at 400$A$~MeV, M4 centrality 
bin, for $Z$=1 particles in three windows of rapidity.}
\label{fig-ad11} 
\end{figure}

\begin{figure}[htb]
\centering\mbox{\epsfig{file=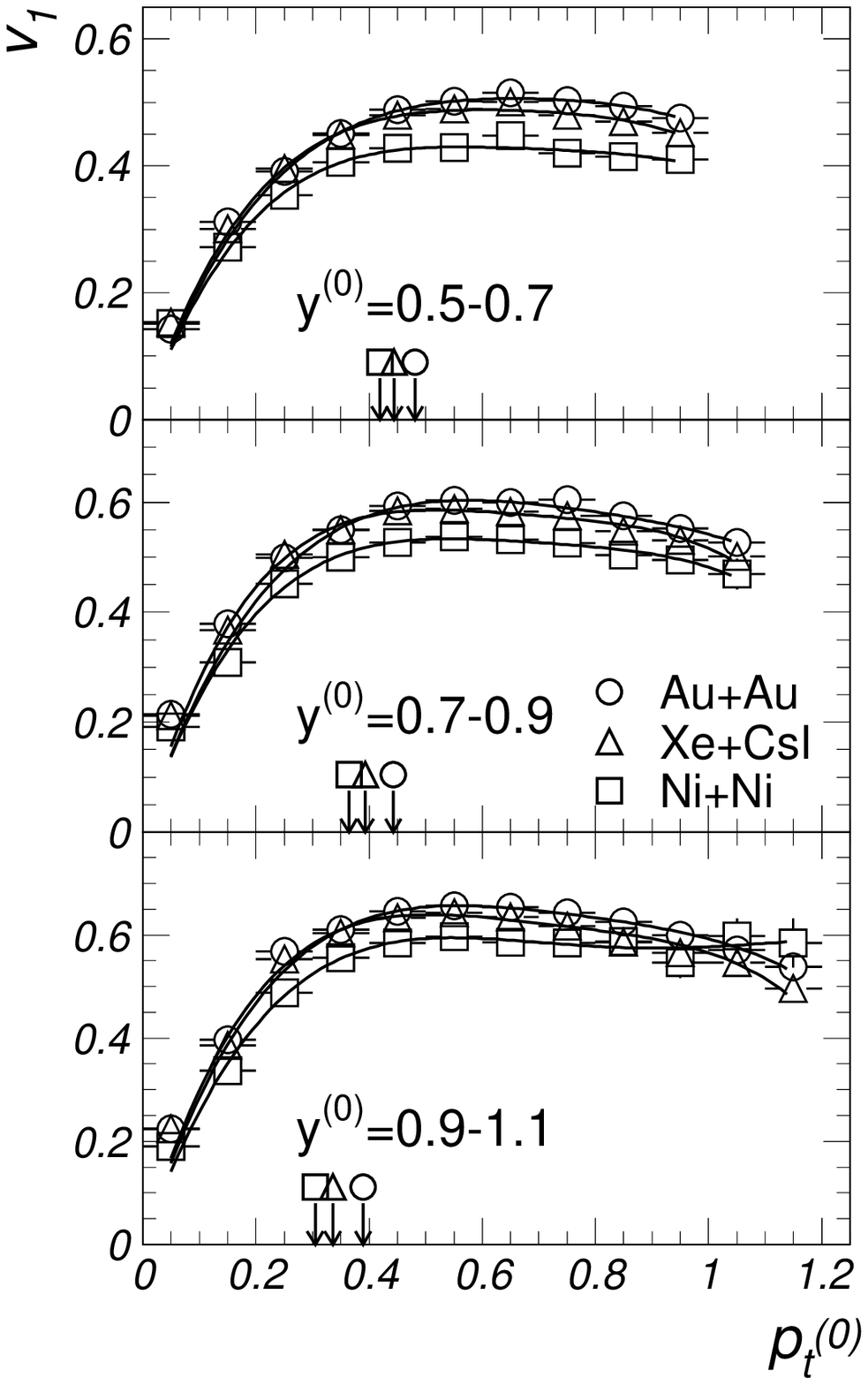, width=.46\textwidth}}
\caption{Differential flow for three systems at 400$A$~MeV, M4 centrality 
bin, for $Z$=2 particles in three windows of rapidity.}
\label{fig-ad12} 
\end{figure}

Similar data as those presented in Fig.~\ref{fig-s2} and \ref{fig-s3} are 
presented in the Appendix for $Z$=1 particles at 250$A$~MeV and for $Z$=1 
and $Z$=2 particles at 400$A$~MeV (Fig.~\ref{fig-ad7} to 
Fig.~\ref{fig-ad12}).

\section{Model Comparison} \label{sect-iqmd}

The IQMD transport model \cite{aic91,har98} is widely used for interpreting 
the data in our energy domain \cite{aic91,par95,cro97b,sof95}.
We use two different parametrizations of the EoS, a hard EoS 
(compressibility $K$=~380 MeV) and a soft EoS ($K$=~200 MeV), 
both with MDI, labeled HM and SM, respectively and without MDI - 
H and S, respectively. 
We use the free nucleon-nucleon cross section, $\sigma_{nn}^{free}$ for all
cases, but for the energy of 90$A$~MeV in addition we consider the case of 
$\sigma_{nn}=0.8 \sigma_{nn}^{free}$.
The events produced by the model are filtered by the experimental filter 
and analyzed in a similar way as the experimental data.
This comprises the same recipe for the centrality selection and the same way 
of reaction plane reconstruction and correction.
However, for the energy of 90$A$~MeV, due to a weak flow signal (see below) 
we prefer to use the true reaction plane for the model calculations.
In what concerns the reaction plane resolution at 400$A$~MeV, it is in the 
model very similar compared to data. For instance, for M4 centrality bin, 
for IQMD SM the correction factors are 1.03, 1.05 and 1.24 for Au, Xe and Ni 
systems, respectively. 

\subsection{What to compare}

A known problem of the IQMD model (and QMD models in general) is that of 
much lower yields of composite fragments compared to data \cite{rei97b}.
For instance, for Au+Au at incident energy of 400$A$~MeV, M4 centrality, 
integrated $Z$=2 yields relative to $Z$=1 are 1/4.7 for the experimental data 
while IQMD predicts 1/25 for HM and 1/15 for SM 
(these ratios do not depend on including MDI or not).

As cluster formation and flow are intimately related, one cannot simply neglect
this dramatic discrepancy.
It is difficult to assess whether this strong dependence of fragment production
on EoS parametrization is a genuine physical effect or is only particular to 
IQMD model.
We note that most of the present models used in heavy-ion collisions at our
energies involve a rather simple phase space coalescence mechanism to produce 
composite particles \cite{ins00,gai01} (IQMD uses the coalescence in coordinate
space only \cite{aic91}).
Efforts to identify the fragments early in the collision might
contribute to clarify this aspect \cite{pur96}.
Promising theoretical candidates to accomplish the task of realistic fragment 
formation could be AMD-type models \cite{ono93}.

\begin{figure}[hbt] 
\centering\mbox{\epsfig{file=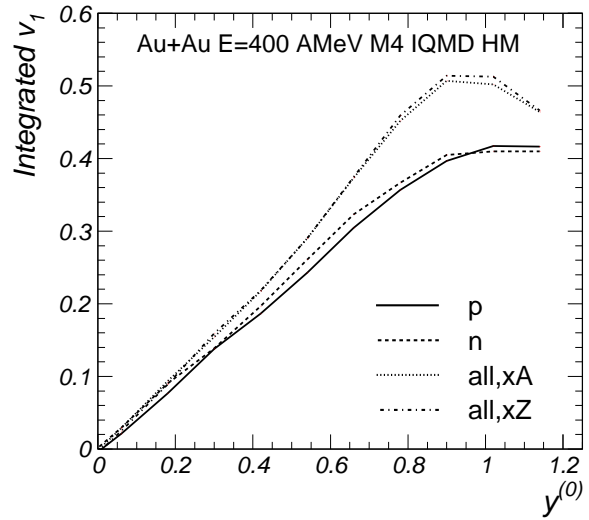,width=.45\textwidth}}
\caption{Comparison of integrated $v_1$ values as a function of rapidity
for protons, neutrons, all particles weighted by mass ($A$) and all particles
weighted by charge ($Z$) for IQMD HM events, for Au+Au at 400$A$~MeV, 
M4 centrality.}
\label{fig-k1} 
\end{figure}

To partially overcome the problem of fragment production in the models,
one can perform the comparison taking into account all charged particles 
weighted by charge $Z$ (so called proton-likes) \cite{ram95,cro97b}.
However, this type of comparison could be biased, as the neutrons bound in
the composite fragments may contribute differently for the calculations 
compared to data.
We have investigated various possibilities by using IQMD events.
The results are presented in Fig.~\ref{fig-k1}, where we compare integrated 
$v_1$ values as a function of rapidity for protons, neutrons, all particles 
weighted by mass ($A$) and all particles weighted by charge ($Z$) for Au+Au 
at 400$A$~MeV, M4 centrality bin.

\begin{figure}[htb] 
\centering\mbox{\epsfig{file=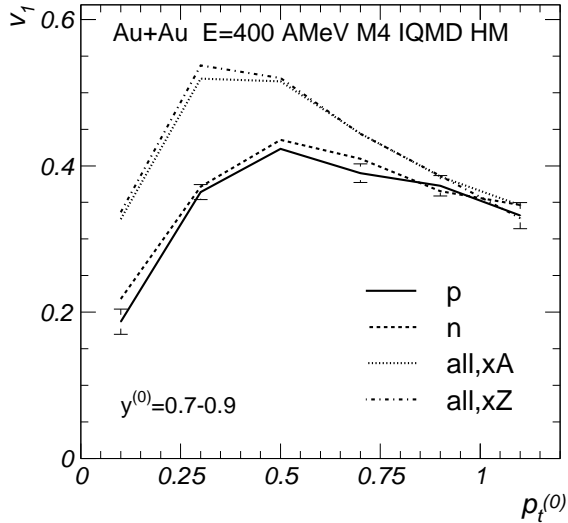,width=.45\textwidth}}
\caption{As Fig.~\ref{fig-k1}, but for differential flow, in the rapidity 
window $y^{(0)}$=0.7-0.9. The statistical errors are plotted for protons.}
\label{fig-k2} 
\end{figure}

In Fig.~\ref{fig-k2} we show the same comparison in case of differential flow.
The neutrons exhibit the same flow as the protons, both for integrated and 
for differential values.
As a consequence, within the model, in both cases the charge-weighted values 
are identical to the mass-weighted ones. However, as fragments heavier than
$Z$=2 are extremely few in the model, this result may be somewhat biased.
In the following we are comparing data and model both for selected particle
types and for proton-likes.

\subsection{Integrated values}

We start our comparison of the data with the IQMD model at the incident energy 
of 90$A$~MeV, for the M4 centrality bin. 
\begin{figure}[hbt] 
\centering\mbox{\epsfig{file=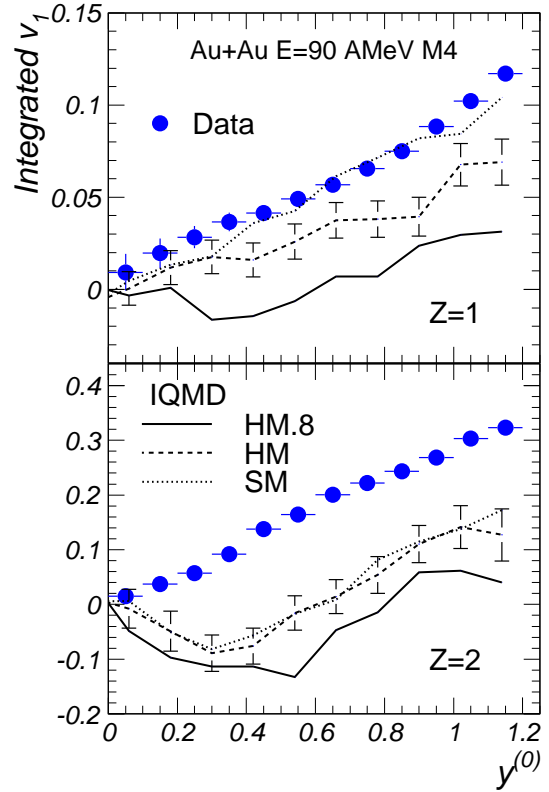, width=.42\textwidth}}
\caption{Integrated $v_1$ values as a function of rapidity for $Z$=1 and $Z$=2 
particles, for the incident energy of 90$A$~MeV. The data points (dots) 
are compared to IQMD calculations for two EoS (lines) parametrizations.
The line labeled HM.8 corresponds to HM case, using 
$\sigma_{nn}=0.8 \sigma_{nn}^{free}$. 
For the HM case the statistical errors of the model are plotted.}
\label{fig-b1} 
\end{figure}

In Fig.~\ref{fig-b1} we show the rapidity dependence of the integrated $v_1$ 
values for particles with $Z$=1 (upper panel) and $Z$=2 (lower panel).
For both species the measured values are compared to the IQMD calculations
for HM and SM parametrizations. For the HM case an additional set of 
calculations has been performed using $\sigma_{nn}=0.8 \sigma_{nn}^{free}$
(labeled HM.8 in Fig.~\ref{fig-b1}).
For the HM case the statistical errors are plotted. For the other cases the
errors are comparable. For data the errors are in most cases smaller than 
the dimension of the points.
The calculated values of the directed flow depend both on the parametrized
EoS and, more pronounced, on $\sigma_{nn}$. 
This dependence is apparently of different magnitude for $Z$=1 and $Z$=2 
particles. 

For the model calculations there is a coexistence of attractive 
(negative $v_1$ values) and repulsive (positive $v_1$) flow, manifested as 
a function of rapidity (we shall call this dual flow). 
This coexistence is different for the two particle species.
We noticed that the above characteristics of the model calculations depend
on centrality as well, both the magnitude of the dual flow and the particle
dependence being enhanced for more peripheral collisions.
The model features are clearly not supported by the data, which show a 
monotonic repulsive flow over all the rapidity domain, both for $Z$=1 and 
$Z$=2 particles, as seen in Fig.~\ref{fig-b1}.
For the experimental data, for the centrality M4 studied here, the reaction 
plane correction factor is 1.54.

\begin{figure}[hbt] 
\centering\mbox{\epsfig{file=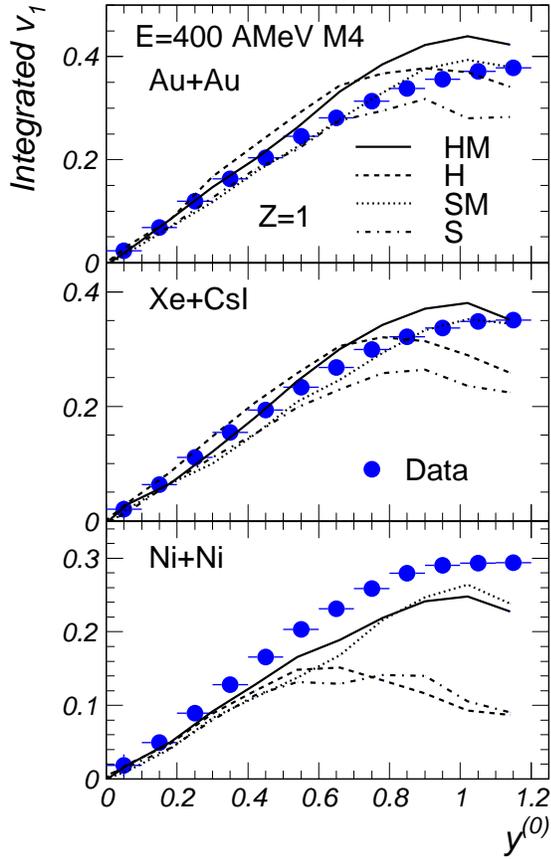, width=.43\textwidth}}
\caption{Integrated $v_1$ values as a function of rapidity, for $Z$=1 
particles, for three systems at 400$A$~MeV, centrality bin M4.
The data points (dots) are compared to IQMD calculations (lines).}
\label{fig-b2} 
\end{figure}

A two-component flow was observed earlier in QMD calculations of
semi-peripheral Ca+Ca collisions at 350$A$~MeV \cite{sof95}. That study 
pointed out its high sensitivity to MDI.
But, unless the discrepancy between the calculations and the measured data 
is resolved, any conclusion on the sensitivity of the directed flow on the 
EoS, $\sigma_{nn}$ or MDI is meaningless for energies around $E_{bal}$.
It is not clear for the moment whether the particle dependence of the dual flow
is not an artifact of the treatment of composite particles in the model.
We note that measurements of $E_{bal}$ for different particle types 
\cite{wes93,mag00} did not reveal, so far, any dual flow.
Calculations with a BUU model \cite{li99} found a dual flow only in a ($p_t$)
differential way, but otherwise monotonic behavior of $\langle p_x\rangle-y$
distributions.
Recent experimental investigations of flow in light systems at $E_{bal}$ 
pointed out interesting aspects of flow of light isotopes and heavy fragments, 
but again the balance energy was found not to depend on particle type 
\cite{cus02}.

\begin{figure}[hbt] 
\centering\mbox{\epsfig{file=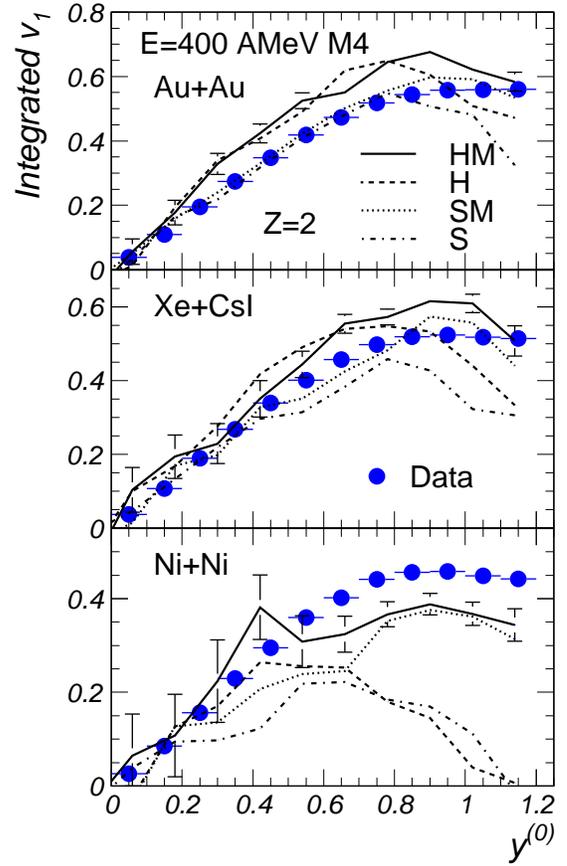, width=.43\textwidth}}
\caption{As Fig.~\ref{fig-b2}, but for $Z$=2 particles.
For the HM case the statistical errors of the model are plotted.}  
\label{fig-b3} 
\end{figure}

In Fig.~\ref{fig-b2} and Fig.~\ref{fig-b3} we show the comparison of data 
and model calculations for $Z$=1 and $Z$=2 particles, respectively.
The three studied systems are considered, for the centrality bin M4.
The integrated $v_1$ values show sensitivity to both EoS and to MDI. 
As expected, the MDI influence the flow essentially in the vicinity 
of the projectile spectator ($y^{(0)}>0.8$). 
This effect is more pronounced the lighter the system.
All these sensitivities are enhanced for $Z$=2 particles.
For Au+Au and Xe+CsI systems, the SM parametrization is reproducing the data 
very well, for both $Z$=1 and $Z$=2 particles. 
This may be the result of a similar balance of thermal and collective 
contributions in the model compared to the data.
In fact, the phase space populations of $Z$=1 and $Z$=2 particles are 
similar for model and data.

In case of Ni+Ni system the dependence on EoS is already negligible,
but obviously the model underestimates the flow.
One parameter of the model, the Gaussian width $L$, which is the phase space 
extension of the wave packet of the particle (and acting as an effective 
interaction range) has been found to influence the directed flow considerably 
\cite{har98}.
A decrease of $L$ for lighter systems has been advocated with the argument of
maximum stability of nucleonic density profiles \cite{har98}.
As no clear prescription exists for handling the value of $L$, we prefer 
to use a constant value of $L$=8.66~fm$^2$ throughout the present work.
A smaller $L$ would lead to an increase of the directed flow \cite{har98} 
and may cure the discrepancy that we observe for the Ni+Ni system, but 
will affect unfavorably the comparison in case of Xe+CsI system.
These effects may reflect the importance played (via the interaction range) 
by the surface. A complete understanding of this aspect is a necessary step 
towards establishing the bulk properties of the nuclear matter created in 
heavy-ion collisions.

We note that comparisons of integrated directed flow for Au+Au system using 
QMD-type models favored mostly a soft EoS \cite{ram95,cro96} but a hard EoS 
was also found to explain another set of experimental data \cite{par95}. 

\begin{figure}[hbt] 
\centering\mbox{\epsfig{file=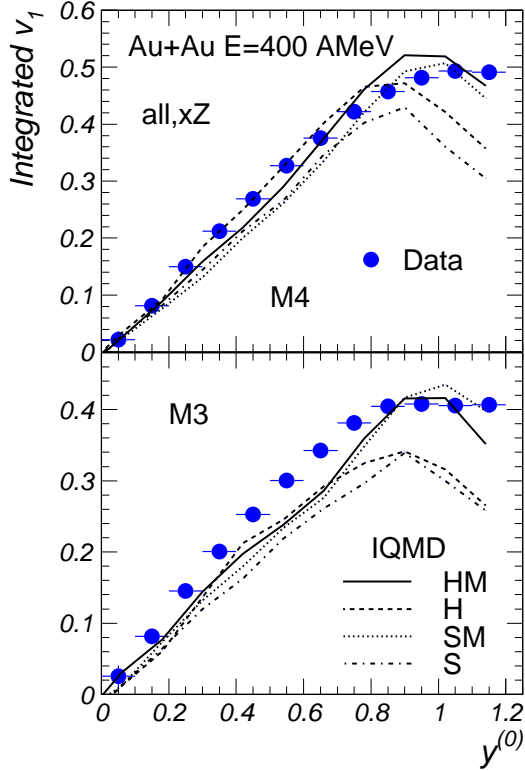, width=.41\textwidth}}
\caption{Integrated $v_1$ values as a function of rapidity for all particles 
weighted by $Z$, for the centrality bins M4 (upper panel) and M3 (lower panel),
for the incident energy of 400$A$~MeV. 
The data points (dots) are compared to IQMD calculations (lines).}
\label{fig-b4} 
\end{figure}

In Fig.~\ref{fig-b4} we present the comparison of the measured integrated 
$v_1$ values to IQMD calculations for Au+Au at the incident energy of 
400$A$~MeV, taking into account all charged particles weighted by charge $Z$.
The centrality bins M4 and M3 are studied.
In this case the sensitivity to EoS is reduced, as a consequence of 
a balance between magnitude of flow and yield of composite particles
in the model: hard EoS produces more flow, but less particles with $Z>$1,
while for soft EoS it is opposite. 
This behavior strongly underlines once more the necessity that
theoretical models appropriately describe the yields of composite 
particles. 
The conclusion on EoS is this time less evident, but 
the parametrizations without MDI are ruled out once again, on the basis of
their departure from the data in the region of spectator rapidity. 
As expected, this effect is more pronounced for the more peripheral centrality
bin M3.

Despite the good agreement seen at the beam energy of 400$A$~MeV, we found 
that in the IQMD model the decrease of flow towards lower incident 
energies is much faster than for data, leading to larger theoretical
$E_{bal}$ compared to data (and to the behavior seen in Fig.~\ref{fig-b1}). 
This may be a result of deficiencies in incorporating MDI and in the treatment
of fragment production. The Pauli blocking may play a role too.
In addition, it has been pointed out that the shape of the flow excitation 
function is drastically influenced by the method of imposing constraints on 
the Fermi momenta \cite{har98}.
The features of the model calculations presented above for 400$A$~MeV
show the danger of deriving EoS-related conclusions 
from rapidity-integrated flow values (like $p_x^{dir}$) 
unless a detailed description of the data is first achieved in a 
differential way. 
As realized early on \cite{aic87}, a soft EoS with MDI is producing
similar magnitude in $p_x^{dir}$ as a hard EoS without MDI. 

\subsection{Differential flow}

We restrict our model comparison of the differential flow to the incident 
energy of 400$A$~MeV and M4 centrality bin. Data for all three systems
investigated so far are compared to the model calculations.

\begin{figure}[hbt]
\centering\mbox{\epsfig{file=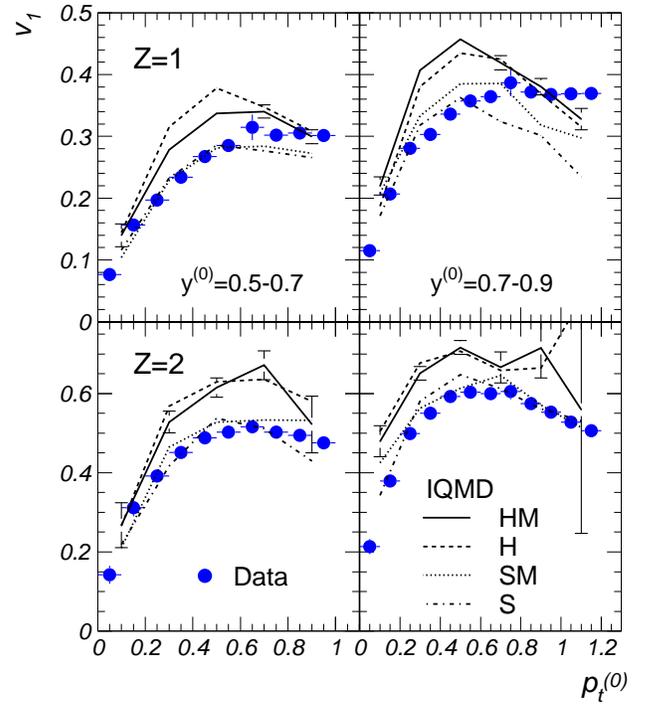, width=.47\textwidth}}
\caption{Differential flow for particles with $Z$=1 (upper row) and $Z$=2 
(lower row) for Au+Au collisions at incident energy of 400$A$~MeV, 
M4 centrality, for two windows in rapidity (columns). 
Experimental data are represented by dots and the model calculations are 
the lines. For the HM case the statistical errors of the model are plotted.} 
\label{fig-d2}
\end{figure} 

In Fig.~\ref{fig-d2} the measured differential directed flow for Au+Au 
collisions at incident energy of 400$A$~MeV, M4 centrality, is compared to 
the IQMD results for all the four parametrizations used above. 
Particles with $Z$=1 (upper row) and $Z$=2 (lower row) for two windows in 
rapidity are used for the comparison.
For both particle species there is a clear sensitivity of DDF on the EoS.
As for the case of the integral flow, the SM parametrization reproduces the
experimental data quite well. 
Apparently the model calculations deviate from the data at high $p_t$ in case 
of $Z$=1 particles, while the corresponding $Z$=2 particles are well explained.
This deviation is more pronounced for larger rapidities. We have found earlier
\cite{cro00} that a BUU model does not explain the DDF of protons in the 
spectator region at higher energies.
The shape of the $Z$=1 DDF distributions are in case of the IQMD model 
strikingly similar to the ones of $Z$=2, while for the data there are subtle 
differences between the two particle species (at this energy of 400$A$~MeV 
as well as down to 90$A$~MeV \cite{and01}).
The model features may be the result that the nucleons (dominating the 
$Z$=1 sample) in the models are all ``primordial'', which does not account 
for the sequential decays of heavier fragments.
The dynamics of the expansion and fragment formation may be responsible for
the differences, too.

\begin{table}[bth]
\caption{Average normalized transverse momentum $\langle p_t^{(0)}\rangle$ 
for particles with $Z$=1 and $Z$=2 in Au+Au collisions at 400$A$~MeV, 
M4 centrality bin.
Data and model values are compared for two rapidity windows.
For data, the number in parenthesis represents the error on the last digit.}
\label{tab-5}
\begin{tabular}{lccc}
Rapidity, particle       &  Data  & IQMD HM  & IQMD SM   \\ \hline
$y^{(0)}$=0.5-0.7 ~~ $Z$=1 &  0.60(3)  & 0.62     & 0.62  \\
~~~~~~~~~~~~~~~~~~~  $Z$=2 &  0.48(2)  & 0.42     & 0.44  \\ \hline
$y^{(0)}$=0.7-0.9 ~~ $Z$=1 &  0.58(3)  & 0.55     & 0.56  \\
~~~~~~~~~~~~~~~~~~~  $Z$=2 &  0.44(2)  & 0.37     & 0.37  \\
\end{tabular}
\end{table}

\begin{figure}[hbt]
\centering\mbox{\epsfig{file=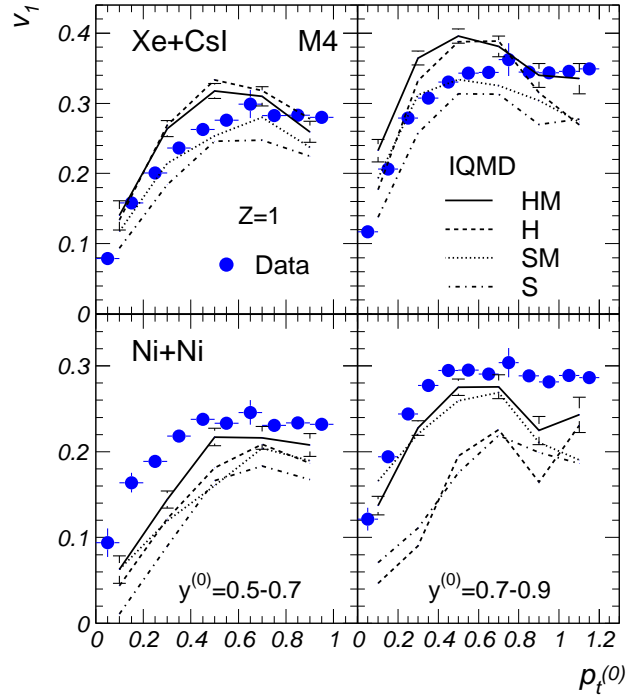, width=.47\textwidth}}
\caption{Comparison of data and model differential flow for Z=1 particles, 
for the incident energy of 400$A$~MeV, M4 centrality bin, for the systems
Xe+CsI (upper row) and Ni+Ni (lower row) for two windows in rapidity
(columns).} 
\label{fig-d3}
\end{figure} 

In Table~\ref{tab-5} we compare the experimental values of the average 
normalized transverse momentum with the values from IQMD, for HM and SM cases.
Particles with $Z$=1 and $Z$=2 for the two windows in rapidity studied in 
Fig.~\ref{fig-d2} are compared. 
The data values have a systematic error represented by the number in 
parenthesis as the error on the last digit.
The model reproduces reasonably well the average transverse momenta for 
$Z$=1 particles, while it underestimates them for $Z$=2, for both 
windows of rapidity. 

We mention that, recently, our experimental differential directed flow 
in Au+Au \cite{and01} was nicely reproduced by a BUU model wich includes 
an improved Dirac-Brueckner formalism \cite{gai01}. In this case, for the 
densities expected at 400$A$~MeV, the EoS is soft, which is in agreement
to our results.

\begin{figure}[hbt]
\centering\mbox{\epsfig{file=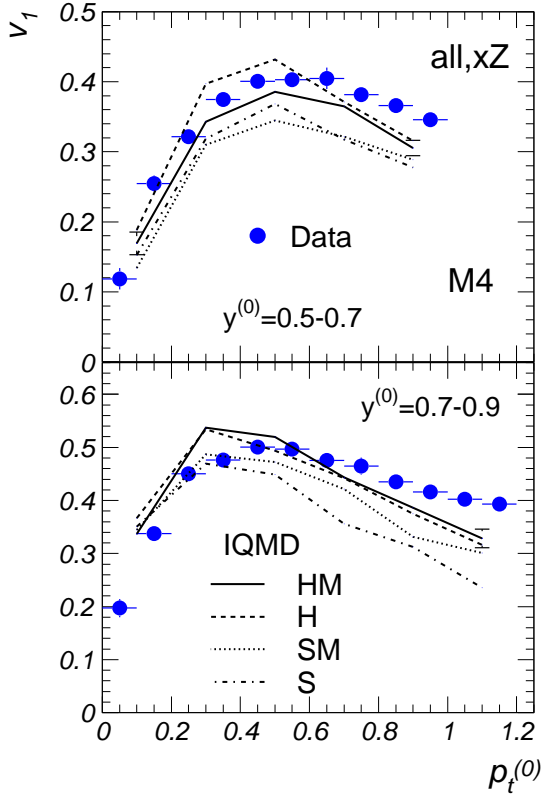, width=.42\textwidth}}
\caption{Differential flow for particles weighted by $Z$ for Au+Au collisions
at incident energy of 400$A$~MeV, M4 centrality, for two windows in rapidity.
Experimental data are represented by dots. The model calculations are the 
lines.} 
\label{fig-d2x}
\end{figure} 

In Fig.~\ref{fig-d3} we show the measured DDF for Xe+CsI and Ni+Ni systems 
at 400$A$~MeV, M4 centrality, in comparison to IQMD results, for $Z$=1 
particles in two windows of rapidity.
In case of Xe+CsI system the model calculations are at the same level of 
agreement with data as in case of Au+Au: the SM parametrization
is reproducing the data, with clear deviations at high momenta.
For the Ni+Ni case even the HM parametrization underpredicts the 
measured data. Most notably, as obvious particularly for Ni+Ni, the MDI have 
effects predominantly at low $p_t$, contrary to earlier BUU predictions 
(performed for the asymmetric system Ar+Pb) \cite{pan93}.

In Fig.~\ref{fig-d2x} we show the model comparison of the differential flow 
for particles weighted by $Z$ for Au+Au collisions at incident energy of 
400$A$~MeV, M4 centrality, for two windows in rapidity. As in case of 
integrated values, as a result of different relative contribution of particles
heavier than $Z$=1, the sensitivity to EoS is reduced for this type of 
comparison. 

\section{Summary and conclusions}

We have presented experimental results on directed flow in Au+Au, Xe+CsI and
Ni+Ni collisions at incident energies from 90 to 400$A$~MeV. 
General features of the directed flow have been investigated using experimental
data, particularly the centrality and the system dependence.
We have studied the rapidity dependence of the first Fourier coefficient, 
$v_1$, integrating over all transverse momentum range.
A special emphasis has been put on the differential directed flow, namely the
$p_t$ dependence of $v_1$.
While for integrated values we presented a new way of looking at old 
(and generally known) dependences, the DDF results are reported for the first
time for our energy domain, both for the centrality and for the system size 
dependence.
We have devoted special care to the corrections of the experimental data.
The influence of the finite granularity of the detector has been studied and
corrected for. The high accuracy of the final results is based as well on a 
good reaction plane resolution achieved with the full coverage of the FOPI 
detector.

We have compared the experimental data with IQMD transport model calculations, 
for both integral and differential $v_1$ values.
This comparison, performed for all the three studied systems,
shows a clear sensitivity of the directed flow on the EoS parametrization 
in the model, especially in case of particle-selected comparison.
In this case, for both integrated and differential directed flow at the 
incident energy of 400$A$~MeV, we conclude that a soft EoS with MDI is the 
only parametrization in the model that reproduces the data for Au and Xe 
systems. A clear discrepancy is seen for Ni system, which needs to be
addressed separately. It may reflect the increasing importance played by 
the nuclear surface for lighter systems.
We consider our present results as a case study on the sensitivities in 
determination of EoS and MDI from directed flow comparisons.
We emphasized the necessity of the present kind of differential comparison
prior to more global quantities.
We have shown that the combination of rapidity and transverse momentum 
analysis of (differential) directed flow can impose constraints on the model.
We also pointed out some difficulties of the model to reproduce the measured 
data concerning:
i) flow at low energy (we considered here 90$A$~MeV),
ii) flow as a function of system size, and
iii) fragment production.
As a consequence, none of the IQMD parametrizations studied here is able to 
consistently explain the whole set of experimental data.

The importance of spectators acting as clocks for the expansion is one
particular argument to study collective flow in semi-central collisions 
at energies from a few hundred MeV to a few GeV per nucleon \cite{dan01}.
We have demonstrated that high precision experimental data allows us to 
study the many facets of the heavy-ion collisions.
Other observables, like $v_2$, should receive a comparable (and simultaneous) 
attention too.
Whether the nuclear equation of state can be extracted from such studies
depends ultimately on the ability of any type of microscopic transport model 
to reproduce the measured features.

\section*{Acknowledgment}
This work has been supported in part by the German BMBF under contracts 
06HD953, RUM-005-95/RUM-99/010, POL-119-95, UNG-021-96 and RUS-676-98 and 
by the Deutsche Forschungsgemeinschaft (DFG) under projects 436 RUM-113/10/0, 
436 RUS-113/143/2 and 446 KOR-113/76/0.
Support has also been received from the Polish State Committee of Scientific 
Research, KBN, from the Hungarian OTKA under grant T029379, 
from the Korea Research Foundation under grant No. KRF-2002-015-CS0009,
from the agreement between GSI and CEA/IN2P3 and from the PROCOPE Program
of DAAD.

\clearpage
\section*{Appendix}

In the following we present additional results that complement those 
included in the body of the paper. As the trends are similar to those
already discussed in section \ref{sect:aa_gen}, we present here only figures.


\end{document}